\documentclass[journal]{IEEEtran}
\usepackage{graphics} 
\usepackage{subfigure}
\usepackage{epsfig} 
\usepackage{multicol}
\usepackage{amssymb}  
\usepackage{mathtools}
\usepackage{amsmath} 
\usepackage{mathrsfs}
\usepackage{amsmath,amsfonts,bm}
\usepackage{algorithm}
\usepackage{algorithmicx}
\usepackage{algpseudocode}
\usepackage{epstopdf}
\usepackage{multirow}
\usepackage{comment}
\usepackage{caption}
\usepackage{soul,xcolor}
\usepackage{hyperref}
\usepackage{array}
\usepackage{balance}
\usepackage{threeparttable}
\newtheorem{theorem}{Theorem}
\newtheorem{remark}{Remark}
\newtheorem{lemma}{Lemma}

\newtheorem{definition}{Definition}

\usepackage{cite}

\ifCLASSINFOpdf
\else
\fi
\usepackage{amsmath}
\usepackage{array}
\begin{document}
%
\title{Distributed Platoon Control Under Quantization: Stability Analysis and Privacy Preservation}
%
%
%
		
\author{
Kaixiang~Zhang, Zhaojian~Li$^{*}$, and Wei~Lin 
\thanks{$*$Zhaojian Li is the corresponding author.}
\thanks{Kaixiang Zhang and Zhaojian Li are with the Department of Mechanical Engineering, Michigan State University, East Lansing, MI 48824, USA (e-mail: \{zhangk64,lizhaoj1\}@msu.edu).}
\thanks{Wei Lin is with the Department of Electrical, Computer, and Systems Engineering, Case Western Reserve University, Cleveland, OH 44106, USA (email: linwei@case.edu). }
}
		
%
%

\markboth{}%
{Shell \MakeLowercase{\textit{et al.}}: Bare Demo of IEEEtran.cls for IEEE Journals}
%



\maketitle
	
\begin{abstract}
Distributed control of connected and automated vehicles has attracted considerable interest for its potential to improve traffic efficiency and safety. However, such control schemes require sharing privacy-sensitive vehicle data, which introduces risks of information leakage and potential malicious activities.
This paper investigates the stability and privacy-preserving properties of distributed platoon control under two types of quantizers: deterministic and probabilistic. For deterministic quantization, we show that the resulting control strategy ensures the system errors remain uniformly ultimately bounded. Moreover, in the absence of auxiliary information, an eavesdropper cannot uniquely infer sensitive vehicle states. In contrast, the use of probabilistic quantization enables asymptotic convergence of the vehicle platoon in expectation with bounded variance. Importantly, probabilistic quantizers can satisfy differential privacy guarantees, thereby preserving privacy even when the eavesdropper possesses arbitrary auxiliary information. We further analyze the trade-off between control performance and privacy by formulating an optimization problem that characterizes the impact of the quantization step on both metrics. Numerical simulations are provided to illustrate the performance differences between the two quantization strategies.
\end{abstract}
	

	%
\IEEEpeerreviewmaketitle
	
\section{Introduction}
Recent developments in wireless communication technologies—particularly vehicle-to-infrastructure (V2I) and vehicle-to-vehicle (V2V) communication—have significantly enhanced the connectivity of modern vehicles, enabling new opportunities for intelligent and coordinated control strategies~\cite{feng2019ARC,Li2017ITSM}. One prominent application is platoon control, which coordinates a group of connected and automated vehicles (CAVs) to travel together as a tightly organized convoy, showing potential for improving traffic flow stability, enhancing roadway safety, and reducing energy consumption~\cite{axelsson2016TITS,Vahidi2018ET,Guo2020ITJ}. The primary objective of platoon control is to ensure that all vehicles in the platoon maintain uniform speed and adhere to the desired inter-vehicle spacing.

From a control systems perspective, a platoon can be modeled as an interconnected system comprising individual vehicle dynamics, inter-vehicle communication topology, spacing policies, and distributed control laws~\cite{Stankovic2000TCST,Kato2002TITS}. The longitudinal dynamics characterize each vehicle's forward motion. When all vehicles share identical dynamics, the system is referred to as homogeneous; otherwise, it is considered heterogeneous~\cite{shladover2015TRR}. Communication protocols determine how vehicles exchange information—what data is shared and with whom—under specific network topologies. The spacing policy defines the target distance between consecutive vehicles and shapes the overall formation structure of the platoon. Each vehicle is equipped with a distributed controller that applies local feedback based on available information, which is typically limited to neighboring vehicles due to sensor and communication range constraints. 
Early research on platoon control dates back to the 1980s, focusing on aspects such as sensing and actuation, control architecture, decentralized implementation, and string stability~\cite{Shladover1991TVT}. 
Since then, significant progress has been made in addressing issues like optimal spacing policies~\cite{Zhou2005TITS,rodonyi2017TITS}, the influence of communication structure~\cite{seiler2004TAC,zheng2015TITS,li2017TCST}, and robustness/adaptation to vehicle system uncertainties~\cite{feng2019TCST,hu2020RAL,viadero2024}. 
More recently, model predictive control methods~\cite{Zheng2017TCST,wang2021TITS,hu2022TITS,qiang2022TITS} have been developed to account for system constraints and improve safety. In parallel, data-driven approaches~\cite{li2021TNNLS,Prathiba2021TVT,chen2024TIV,liu2022ITJ} such as reinforcement learning and dynamic programming have emerged as promising alternatives to model-based control by leveraging real-time data to guide controller design.

While distributed platoon control enables efficient coordination among CAVs, it also introduces significant privacy concerns. Achieving cooperative behavior requires extensive sharing of onboard vehicle data, which often contains sensitive or private information, through V2V communication. In a typical distributed control framework, each vehicle transmits its measured or estimated states to its neighbors, then computes and applies a local control action based on the received data. This continuous exchange of information across the network exposes system measurements to potential interception, making the communication channels vulnerable to eavesdropping. An external eavesdropper could exploit this vulnerability to infer private vehicle data. Prior studies have demonstrated that exposing internal vehicle information through networked communication can lead to various security threats and malicious behaviors~\cite{PetitTITS2015,Amoozadeh2015CM,Sun2022TITS}. Without effective privacy protection, such breaches could result in severe consequences for CAVs and other vehicles sharing the roadway.

Given the rising importance of cybersecurity in intelligent vehicle systems, ensuring the privacy of CAVs in distributed platoon control has become a critical concern. Although privacy and security issues have been extensively explored in various intelligent transportation scenarios~\cite{zhang2024TITS,Farivar2021TITS,Biroon2020ACC,Gao2022TITS}, protecting sensitive information during inter-vehicle communication remains particularly challenging in the context of real-time, resource-constrained platooning systems. Existing privacy-preserving strategies can be broadly categorized into encryption-based~\cite{pan2023TSMCS,he2024TITS} and perturbation-based~\cite{chen2025SJ} approaches. Encryption techniques rely on cryptographic algorithms to conceal sensitive data, offering strong privacy guarantees. However, their high computational overhead and latency often make them unsuitable for embedded systems with limited onboard processing capabilities. In contrast, perturbation-based methods inject deliberate noise, such as random or uncorrelated signals, into the transmitted data to obscure the true system states. While computationally efficient, these methods inherently involve a trade-off between control performance and privacy, as excessive noise can degrade system stability and responsiveness.
Recently, quantization has emerged as a lightweight yet effective alternative for privacy protection in areas such as distributed optimization~\cite{wang2022TAC}, networked control~\cite{liu2023CDC}, and machine learning~\cite{youn2023arXiv}. Although quantization has been employed in distributed platoon control to reduce communication load or to examine its impact on control performance~\cite{zhu2023TVT,cui2022TVT,zhao2024TITS}, its potential for privacy preservation remains underexplored. Like perturbation-based techniques, quantization introduces structured noise into the system, which can obscure sensitive information but may also affect control quality. This dual effect highlights the need to systematically investigate how different types of quantizers (e.g., deterministic and probabilistic ones) influence both the stability and the privacy of distributed platoon systems. Understanding this relationship is essential for designing quantization strategies that strike a desirable balance between secure communication and reliable control performance.

This paper explores the stability and privacy-preserving characteristics of distributed platoon control under both deterministic and probabilistic quantization schemes. Rather than transmitting exact vehicle state information, each  vehicle applies quantization to obscure its true states before sharing data across the communication network. For the deterministic case, the corresponding distributed control strategy guarantees uniform ultimate boundedness of the system errors. To assess privacy, we extend the concept of $l$-diversity~\cite{machanavajjhala2007ACM}, showing that when an eavesdropper lacks auxiliary knowledge of the system, it cannot uniquely infer the original vehicle states from the quantized data.
In the case of probabilistic quantization, we prove that the system achieves asymptotic convergence in expectation, with the error variance bounded by a value that depends on the quantization step. Furthermore, we establish that the probabilistic quantizer enables differential privacy~\cite{Dwork2014FTTCS,cortes2016CDC}, a widely adopted standard that offers strong protection even when adversaries possess arbitrary auxiliary information. 
Since both control performance and privacy guarantees are influenced by the quantization step, an optimization problem is formulated to explicitly characterize the trade-off between these competing objectives.

The main contributions of the paper are as follows: First, different from existing works~\cite{zhu2023TVT,cui2022TVT,zhao2024TITS} that focus solely on the impact of quantization on control performance or communication efficiency, this paper presents a comprehensive study on the stability and privacy-preserving properties of distributed platoon control under the deterministic and probabilistic quantization schemes. Our findings reveal that quantization can serve not only as a tool for efficient communication but also as a lightweight and practical mechanism for privacy protection in real-time, resource-constrained CAV applications. Second, to the best of our knowledge, this is the first time that the probabilistic quantization is incorporated into the distributed platoon control. We prove convergence of system errors in the mean sense with bounded variance and show that the rigorous differential privacy can be achieved. Finally, extensive simulations are conducted to evaluate and compare the performance of the two quantization schemes. The results demonstrate that compared to its deterministic counterpart, the probabilistic quantizer achieves superior control performance while guaranteeing stronger privacy preservation when the eavesdropper has access to full auxiliary information of the platoon system.

The remainder of the paper is organized as follows. Section~\ref{sec_formulation} introduces the necessary notations and formulates the distributed platoon control problem. Section~\ref{sec_deterministic} analyzes the stability and privacy-preserving properties of the deterministic quantizer. Section~\ref{sec_probabilistic} investigates the convergence behavior and differential privacy guarantees of the probabilistic quantizer. Simulation results are presented in Section~\ref{sec_perfEva} to evaluate the performance of both schemes. Finally, Section~\ref{sec_conclusion} concludes the paper.
	
\textit{Notations:} We denote $\mathbb{R}$ and $\mathbb{Z}$ as the set of real numbers and integers, respectively. 
Let $\lambda_{i}(A)$ denote the $i$-th eigenvalue of matrix $A\in \mathbb{R}^{n\times n}$, $i=1, 2, \cdots, n$, and the eigenvalues are represented in an increasing order based on their real parts. $\lambda_{\mathrm{max}}(A)$ ($\lambda_{\mathrm{min}}(A)$) denotes the maximum (minimum) eigenvalue of matrix $A$.
Let $1_n$ denote an $n\times 1$ vector with all entries being ones, and $I_{n}$ denote an $n\times n$ identity matrix. 
The notation $\mathrm{diag}(a_{1}, a_{2}, \ldots, a_{n})$ represents a diagonal matrix whose diagonal entries are $a_{1}, a_{2}, \ldots, a_{n}$.
The symbol $\otimes$ denotes the Kronecker product.
	
\section{Modeling and Problem Description} \label{sec_formulation}

\subsection{Communication Topology} \label{subsec_topology}
\begin{figure}[!t]
	\setlength{\abovecaptionskip}{0pt}
	\centering
	\subfigure[] {\label{fig:vehicle_platoon}
		\includegraphics[width=0.45\textwidth]{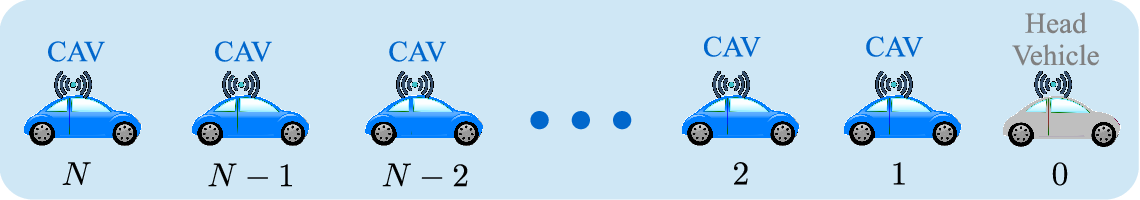}
	}
	\subfigure[] {\label{fig:BD}
		\includegraphics[width=0.4115\textwidth]{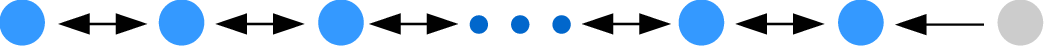}
	}
	\subfigure[] {\label{fig:BDL}
		\includegraphics[width=0.4115\textwidth]{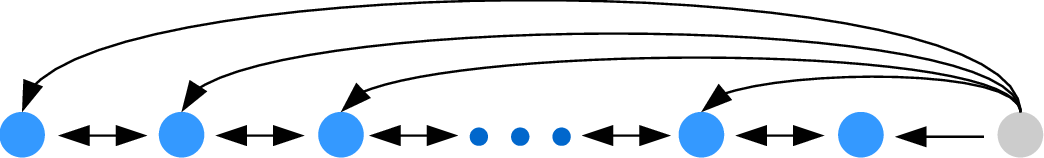}
	}
	\subfigure[] {\label{fig:PF}
		\includegraphics[width=0.4115\textwidth]{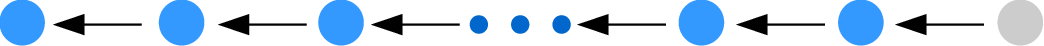}
	}
    \subfigure[] {\label{fig:PLF}
	    \includegraphics[width=0.4115\textwidth]{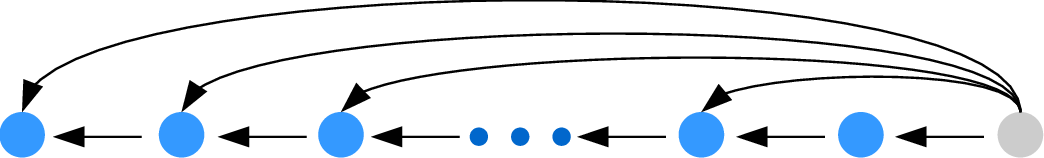}
    }
    \subfigure[] {\label{fig:TPF}
    	\includegraphics[width=0.4115\textwidth]{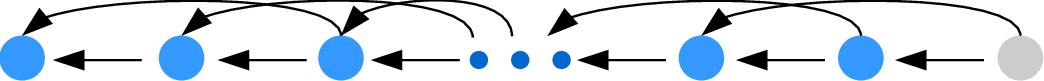}
    }
    \subfigure[] {\label{fig:TPLF}
    	\includegraphics[width=0.4115\textwidth]{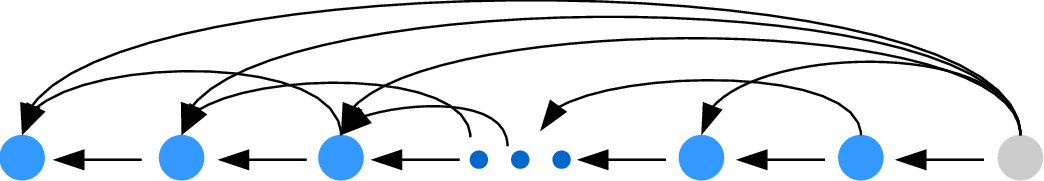}
    }
	\caption{Schematic of platoon system and communication topology. (a) Platoon structure with $N+1$ vehicles. Typical communication topologies: (b) BD, (c) BDL, (d) PF, (e) PLF, (f) TPF, and (g) TPLF.}
	\label{fig:SystemSchematic}
\end{figure}

As illustrated in Fig.~\ref{fig:SystemSchematic}, the considered platoon system consists of $N+1$ vehicles: one head vehicle (indexed as $0$) and $N$ following vehicles (indexed from 1 to $N$). The V2V communication flow among the followers is modeled by a directed graph $\mathcal{G}=\{\mathcal{V}, \mathcal{E} \}$ with the node set $\mathcal{V}=\{1, 2, \cdots, N\}$ and the edge set $\mathcal{E}\subset \mathcal{V}\times \mathcal{V}$. 
A directed edge $(i, j)\in \mathcal{E}$ indicates that vehicle $i$ can receive information from vehicle $j$, and vehicle $j$ is said to be a neighbor of vehicle $i$.
The adjacent matrix associate with graph $\mathcal{G}$ is denoted by $M=[m_{ij}]\in \mathbb{R}^{N\times N}$, where $m_{ij}$ is defined as
\[
\begin{cases}
	m_{ij} = 1, & \text{if}\, (i, j)\in \mathcal{E},
	\\
	m_{ij} = 0, & \text{if}\, (i, j)\notin \mathcal{E}.
\end{cases}
\] 
The corresponding Laplacian matrix $L=[l_{ij}]\in \mathbb{R}^{N\times N}$ is defined as
\[
l_{ij} = \begin{cases}
	-m_{ij}, & i\neq j,
	\\
	\sum_{k=1}^{N}m_{ik}, & i= j.
\end{cases}
\]
Furthermore, communication from the head vehicle to the following vehicles is described by a diagonal pinning matrix $S=\mathrm{diag}\{s_{1}, s_{2}, \cdots, s_{N}\}$, where $s_{i}=1$ if vehicle $i$ can directly receive information from the head vehicle, and $s_{i}=0$ otherwise.   

The communication topology in this paper satisfies two mild but essential conditions: $\left. 1 \right)$ At least one of the following vehicles can receive information from the head vehicle, and there exists a (not necessarily unique) directed path from the head vehicle to every following vehicle. This implies that all followers are indirectly or directly connected to the leader. $\left. 2 \right)$ The matrix $L+S$ has real and strictly positive eigenvalues, i.e., $0<\lambda_{1}(L+S)\le \lambda_{2}(L+S)\le \cdots \le \lambda_{N}(L+S)$. These requirements are commonly adopted in distributed platoon control. Fig.~\ref{fig:SystemSchematic} shows six representative topologies satisfying these conditions: bidirectional (BD) topology, bidirectional-leader (BDL) topology, predecessor following (PF) topology, predecessor-leader following (PLF) topology, two-predecessors following (TPF) topology, and two-predecessor-leader following (TPLF) topology. 
For brevity, topologies with complex eigenvalues are omitted; however, the proposed methods and theoretical results can be extended to such cases following similar techniques. 

\subsection{Vehicle Longitudinal Dynamics}
The platoon is modeled as a group of interconnected nodes, each representing a vehicle. The longitudinal dynamics of each vehicle include effects from the engine, braking, and aerodynamic drag. Based on standard modeling assumptions~\cite{hu2020RAL,Ghasemi2013TVT}, the dynamics of vehicle $i$ are give by 
\begin{equation} \label{equ:longitudinal}
\begin{cases}
	\dot{p}_{i} = v_{i},
	\\
	\dot{v}_{i} = a_{i},
	\\
	\dot{a}_{i} = f_{i}(v_{i}, a_{i}) + \frac{b_{i}}{\tau_{i} m_{i}},
\end{cases} i= 1, 2, \cdots, N,
\end{equation}
where $p_{i}(t)$, $v_{i}(t)$, and $a_{i}(t)$ represent the position, velocity, and acceleration of vehicle $i$, $b_{i}(t)$ is the engine input, $m_{i}$ is the vehicle mass, and $\tau_{i}$ denotes the inertial delay. The nonlinear term $f_{i}(v_{i}, a_{i})$ is defined as 
\[
\begin{aligned}
f_{i}(v_{i}, a_{i}) = &-\frac{1}{\tau_{i}}\left( a_{i} + \frac{\sigma \phi_{i} c_{di}}{2m_{i}}v_{i}^{2}(t) + \frac{d_{mi}}{m_{i}} \right) 
\\
&
-\frac{\sigma \phi_{i} c_{di}}{m_{i}} v_{i}(t)a_{i},
\end{aligned}
\]
where $\sigma$ is the specific mass of the air, $\phi_{i}$ is the cross-sectional area, $c_{di}$ denotes the drag coefficient, and $d_{mi}$ is the mechanical drag. To transform the nonlinear model~\eqref{equ:longitudinal} into a linear one, $b_{i}(t)$ is designed as
\begin{equation} \label{equ:bi}
	b_{i} = m_{i}u_{i} + \frac{\sigma \phi_{i} c_{di}}{2}v_{i}^{2} + d_{mi} + \sigma \phi_{i} c_{di} v_{i}a_{i},
\end{equation}
with $u_{i}(t)$ being the new control input. After substituting~\eqref{equ:bi} into \eqref{equ:longitudinal}, the linear model for vehicle longitudinal dynamics is obtained, as follows:
\begin{equation} \label{equ:LTI}
	\dot{x}_{i} = A_{i}x_{i} + B_{i}u_{i},
\end{equation}
where 
\[
x_{i}(t) = \begin{bmatrix}
	p_{i}(t) \\ v_{i}(t) \\ a_{i}(t)
\end{bmatrix}, \quad A_{i} = \begin{bmatrix}
0 & 1 & 0 \\ 0 & 0 & 1 \\ 0 & 0 & -\frac{1}{\tau_{i}} \end{bmatrix}, \quad B_{i} = \begin{bmatrix}
	0 \\ 0 \\ \frac{1}{\tau_{i}}
\end{bmatrix}.
\]
In this paper, it is assumed that the platoon is homogeneous, i.e., $A_{i}=A$ and $B_{i}=B$ for all $i=1, 2, \cdots, N$. The system state of the head vehicle is similarly defined as $x_{0}(t) = \begin{bmatrix}
	p_{0}(t), v_{0}(t), a_{0}(t)
\end{bmatrix}^{\top}$, where $p_{0}(t)$, $v_{0}(t)$, and $a_{0}(t)$ denote the position, velocity, and acceleration of the head vehicle. At steady state, the head vehicle is considered to be of constant-velocity type, i.e., $p_{0}=v_{0}t$ and $a_{0}=0$.

\subsection{Problem Formulation} \label{subsec_problem}
The objective of platoon control is to ensure that all the following vehicles track the speed of the head vehicle while maintaining a constant inter-vehicular distance. Specifically, let $d_{r}$ be the desired constant distance between two consecutive vehicles. The control objective then can be formulated as
\begin{equation} \label{equ:obj}
\begin{cases}
	\lim_{t\rightarrow \infty} p_{0}(t)-p_{i}(t) = id_{r},
	\\
	\lim_{t\rightarrow \infty} v_{0}(t)-v_{i}(t) = 0,
	\\
	\lim_{t\rightarrow \infty} a_{0}(t)-a_{i}(t) = 0,
\end{cases} i=1, 2, \cdots, N.
\end{equation} 
According to~\eqref{equ:obj}, the tracking error $\varepsilon_{i}(t)$ for each following vehicle is defined as
\begin{equation} \label{equ:varepsilon}
	\varepsilon_{i} = x_{i}+d_{i}-x_{0},
\end{equation}
where $d_{i}=\begin{bmatrix}
	id_{r}, 0, 0
\end{bmatrix}^{\top}$. Based on the definition of $A$ and $d_{i}$, it is easy to verify that $Ad_{i} = 0$ and $\dot{d}_{i}=0$. Given the head vehicle runs at a constant velocity, we have $\dot{x}_{0}(t) = Ax_{0}(t)$. Using~\eqref{equ:LTI}, \eqref{equ:varepsilon}, and the aforementioned properties, it can be concluded that
\begin{equation} \label{equ:LTI_varepsilon}
	\dot{\varepsilon}_{i} = A\varepsilon_{i} + Bu_{i}.
\end{equation} 

To ensure $\lim_{t\rightarrow \infty} \varepsilon_{i}(t)=0$, the following distributed controller can be applied to each vehicle:
\begin{equation} \label{equ:DC}
\begin{aligned}
	u_{i} = & K\left( \sum_{j=1}^{N} m_{ij}\left( (x_{j}+d_{j}) - (x_{i}+d_{i}) \right) \right.
	\\
	& \qquad \qquad \qquad \qquad \left. + s_{i}\left( x_{0} - (x_{i}+d_{i}) \right) \right),
\end{aligned}
\end{equation}
where 
$K=B^{\top}P$, and $P>0$ is a positive definite matrix that satisfies 
\begin{equation} \label{equ:P}
	PA+A^{\top}P-2\lambda_{1}(L+S)PBB^{\top}P+\gamma I_{3}\le 0
\end{equation}  
for some $\gamma>0$.

To implement the distributed control scheme in~\eqref{equ:DC}, the head vehicle needs to broadcast $x_{0}(t)$ to its connected followers, and each following vehicle should transmit its state $x_{i}(t)$ to its neighbors. However, this shared data may include privacy-sensitive information that can be exploited by eavesdroppers.
In this work, we focus on the following attack model~\cite{Gao2022TITS}:
\begin{itemize}
	\item \emph{Eavesdropping attacks:} An external eavesdropper intercepts V2V communications to access transmitted messages, intending to extract private information about the transmitting parties.
\end{itemize} 

Specifically, we assume that the states of the involved vehicles, i.e., $x_{0}(t), x_{1}(t), x_{2}(t), \cdots, x_{N}(t)$, contain privacy-sensitive information. Under the control framework in~\eqref{equ:DC}, an external eavesdropper can successfully wiretap the messages $x(k)$. To mitigate this risk, this paper applies quantization techniques to conceal the information exchanged in the vehicle communication network. In particular, our aim is to study how deterministic and probabilistic quantization affect the stability and privacy-preserving properties of the distributed platoon control system.

\section{Deterministic Quantization} \label{sec_deterministic}
In this section, we develop a distributed control law based on deterministic quantization and analyze the resulting stability and privacy properties. We first define the quantizer and then design a quantized control strategy to ensure uniform ultimate boundedness of the system errors. Finally, we assess the privacy protection offered by the deterministic quantizer.

\subsection{Deterministic Quantizer for Platoon Control} \label{subsec_det_quantizer}
To protect sensitive vehicle state information, each vehicle applies a deterministic quantizer to mask its data before sharing it with neighbors. Given a vector $z=\begin{bmatrix}
	z_{1}, z_{2}, \cdots, z_{m}
\end{bmatrix}^{\top}\in \mathbb{R}^{m}$, the deterministic quantizer is defined as $\mathcal{Q}_{d}(z) = \begin{bmatrix}
\mathcal{Q}_{d}(z_{1}), \mathcal{Q}_{d}(z_{2}), \cdots, \mathcal{Q}_{d}(z_{m})
\end{bmatrix}^{\top}$, where each component $\mathcal{Q}_{d}(z_{\ell})$  for $\ell=1,2, \cdots, m$ is given by
\begin{equation} \label{equ:det_quantizer}
\begin{aligned}
	\mathcal{Q}_{d}(z_{\ell}) &= \begin{cases}
		n\Delta, & z_{\ell}-n\Delta < (n+1)\Delta-z_{\ell}, \\
		(n+1)\Delta, & z_{\ell}-n\Delta \ge (n+1)\Delta-z_{\ell}, 
	\end{cases}
    \\
    & \qquad \qquad \qquad \quad \,\, z_{\ell}\in \left(n\Delta, (n+1)\Delta\right], n\in \mathbb{Z}, 
\end{aligned}
\end{equation}
and $\Delta>0$ denotes the quantization step. From~\eqref{equ:det_quantizer}, it follows that the quantization error satisfies $| \mathcal{Q}_{d}(z_{\ell}) - z_{\ell} | \le \frac{\Delta}{2}$.
The deterministic quantizer maps a continuous input to a discrete output level using a fixed rounding rule. Thus, for any given input, the output of the deterministic quantizer is always the same, making the quantization process predictable. 

Substituting the quantized data into the distributed controller~\eqref{equ:DC} yields the modified control law:
\begin{equation} \label{equ:DC_deterministic}
\begin{aligned}
	u_{i} = & K\left( \sum_{j=1}^{N} m_{ij}\left( (\mathcal{Q}_{d}(x_{j})+d_{j}) - (\mathcal{Q}_{d}(x_{i})+d_{i}) \right) \right.
	\\
	& \qquad \qquad \qquad \left. + s_{i}\left( \mathcal{Q}_{d}(x_{0}) - (\mathcal{Q}_{d}(x_{i})+d_{i}) \right) \right).
\end{aligned}
\end{equation}
To facilitate the following analysis, define the quantization errors as
\begin{equation} \label{equ:e_det}
\begin{aligned}
	e_{d0} &= \mathcal{Q}_{d}(x_{0})-x_{0}, 
	\\
	e_{di} &= \mathcal{Q}_{d}(x_{i})-x_{i},\; i=1, 2, \cdots, N.
\end{aligned}
\end{equation}
After substituting~\eqref{equ:DC_deterministic} and~\eqref{equ:e_det} into~\eqref{equ:LTI_varepsilon} and using $\varepsilon_{j}(t)-\varepsilon_{i}(t) = (x_{j}(t)+d_{j})-(x_{i}(t)+d_{i})$, the closed-loop dynamics of vehicle $i$ can be derived, as follows:
\begin{equation} \label{equ:dot_varepsilon_i}
\begin{aligned}
	\dot{\varepsilon}_{i} = A\varepsilon_{i} &- BK\left( \sum_{j=1}^{N} m_{ij}\left( \varepsilon_{i} - \varepsilon_{j} \right) 
	+ s_{i}\varepsilon_{i} \right)
	\\
	&- BK\left( \sum_{j=1}^{N} m_{ij}\left( e_{di}-e_{dj} \right) 
	+ s_{i}\left(e_{di}-e_{d0} \right) \right).
\end{aligned}
\end{equation}
The collective tracking errors of all following vehicles are defined as
\begin{equation}
	\varepsilon = \begin{bmatrix} \varepsilon_{1}^{\top}, \varepsilon_{2}^{\top}, \cdots, \varepsilon_{N}^{\top} \end{bmatrix}^{\top}.
\end{equation}
Based on~\eqref{equ:dot_varepsilon_i} and the communication topology introduced in Section~\ref{subsec_topology}, the overall closed-loop dynamics of the homogeneous platoon can be expressed in the following compact form:
\begin{equation} \label{equ:dot_varepsilon}
\begin{aligned}
	\dot{\varepsilon} &= \left( I_{N}\otimes A - (L+S)\otimes BK \right)\varepsilon 
	\\
	&\qquad \qquad - \left((L+S)\otimes BK \right)e_{d} + \left( S\otimes BK \right) \left(1_{N}\otimes e_{d0} \right)
	\\
	&= \left( I_{N}\otimes A - (L+S)\otimes BK \right)\varepsilon 
	\\
	&\qquad \qquad - \left((L+S)\otimes BK \right)\left( e_{d}-1_{N}\otimes e_{d0} \right),
\end{aligned}
\end{equation}
where $e_{d}(t) = \begin{bmatrix}
	e_{d1}^{\top}(t), e_{d2}^{\top}(t), \cdots, e_{dN}^{\top}(t)
\end{bmatrix}^{\top}$, and the second equality is derived by using the property $L 1_{N}=0$.

\begin{theorem}
	Under the deterministic quantization scheme, the distributed platoon controller~\eqref{equ:DC_deterministic} ensures that the collective tracking error $\varepsilon(t)$ is uniformly ultimately bounded (UUB).
\end{theorem}

\begin{IEEEproof}
There exists a nonsingular matrix $U\in \mathbb{R}^{N\times N}$ such that
\begin{equation} \label{equ:Lambda}
	L+S = U\Lambda U^{-1},
\end{equation}
where $\Lambda \in \mathbb{R}^{N\times N}$ is the Jordan normal form of $L+S$, and its diagonal entries are the eigenvalues $\lambda_{i}(L+S)$. Define a transformed error variable
\begin{equation} \label{equ:tilde_varepsilon}
	\tilde{\varepsilon} = \left( U^{-1}\otimes I_{3} \right) \varepsilon.
\end{equation}
From~\eqref{equ:dot_varepsilon}-\eqref{equ:tilde_varepsilon}, we have 
\begin{equation} \label{equ:dot_tilde_varepsilon}
\begin{aligned}
	\dot{\tilde{\varepsilon}} &= \left( U^{-1}\otimes I_{3} \right) \dot{\varepsilon}
	\\
	&= \left( I_{N}\otimes A - \Lambda\otimes BK \right) \tilde{\varepsilon} 
	\\
	&\qquad \qquad - \left(\Lambda\otimes BK \right) \left( U^{-1}\otimes I_{3} \right) \left( e_{d}-1_{N}\otimes e_{d0} \right).
\end{aligned}
\end{equation}

A Lyapunov function $V(t)\in \mathbb{R}$ is designed as 
\begin{equation} \label{equ:V}
	V = \tilde{\varepsilon}^{\top}(I_{N}\otimes P) \tilde{\varepsilon},
\end{equation}
where $P>0$ satisfies the condition in~\eqref{equ:P}. Based on~\eqref{equ:dot_tilde_varepsilon} and $K=B^{\top}P$, the time derivative of $V(t)$ can be obtained, as follows:
\begin{equation}
\begin{aligned}
	\dot{V} &= \dot{\tilde{\varepsilon}}^{\top}(I_{N}\otimes P) \tilde{\varepsilon} + \tilde{\varepsilon}^{\top}(I_{N}\otimes P) \dot{\tilde{\varepsilon}}
	\\
	&= \tilde{\varepsilon}^{\top}(I_{N}\otimes (PA+A^{\top}P)-(\Lambda+\Lambda^{\top})\otimes PBB^{\top}P ) \tilde{\varepsilon}
	\\
	&\quad - 2\tilde{\varepsilon}^{\top} \left(\Lambda\otimes PBB^{\top}P \right) \left( U^{-1}\otimes I_{3} \right) \left( e_{d}-1_{N}\otimes e_{d0} \right).
\end{aligned}
\end{equation}
From~\eqref{equ:P} and $\lambda_{1}(L+S)\le \lambda_{2}(L+S)\le \cdots \le \lambda_{N}(L+S)$, it can be concluded that for all $i=1, 2, \cdots, N$, $PA+A^{\top}P-2\lambda_{i}(L+S)PBB^{\top}P+\gamma I_{3}\le 0$, which indicates that
\begin{equation} \label{equ:cond1}
	I_{N}\otimes (PA+A^{\top}P)-(\Lambda+\Lambda^{\top})\otimes PBB^{\top}P \le -\gamma I_{3N}.
\end{equation}
In addition, as $e_{d0}(t)$ and $e_{d}(t)$ are the errors induced by the deterministic quantizer, we have 
\begin{equation} \label{equ:cond2}
    \left\| e_{d}-1_{N}\otimes e_{d0} \right\| \le \sqrt{3N}\Delta.
\end{equation}
With~\eqref{equ:cond1} and \eqref{equ:cond2}, $\dot{V}(t)$ can be upper bounded by
\begin{equation} \label{equ:bound}
\begin{aligned}
	\dot{V} &\le -\gamma \tilde{\varepsilon}^{\top}\tilde{\varepsilon} + 2 \| \tilde{\varepsilon} \|  \|\Lambda\| \|PBB^{\top}P \| \| U^{-1}\| \| e_{d}-1_{N}\otimes e_{d0} \|
	\\
    &\le -\gamma \tilde{\varepsilon}^{\top}\tilde{\varepsilon} + 2\sqrt{3N} \lambda_{N}(L+S) \| \tilde{\varepsilon} \| \|PBB^{\top}P \| \| U^{-1}\| \Delta.
\end{aligned}
\end{equation}
Invoking Theorem 4.18 from~\cite{khalil2002nonlinear}, we conclude that $\tilde{\varepsilon}(t)$ is UUB and $\lim_{t\rightarrow \infty} \| \tilde{\varepsilon}(t) \| \le \sqrt{\frac{\lambda_{\mathrm{max}}(P)}{\lambda_{\mathrm{min}}(P)}}  \frac{2\sqrt{3N} \lambda_{N}(L+S) \|PBB^{\top}P \| \| U^{-1}\| \Delta}{\gamma}$. From~\eqref{equ:tilde_varepsilon}, it can be further obtained that $\varepsilon(t)$ is UUB and $\lim_{t\rightarrow \infty} \| \varepsilon(t) \| \le \sqrt{\frac{\lambda_{\mathrm{max}}(P)}{\lambda_{\mathrm{min}}(P)}}  \frac{2\sqrt{3N} \lambda_{N}(L+S) \|PBB^{\top}P \| \Delta}{\gamma}$. 
\end{IEEEproof}

\subsection{Privacy Analysis} \label{subsec_diversity}
We now analyze the privacy guarantees provided by deterministic quantization. As discussed in Section~\ref{subsec_problem}, the external eavesdropper seeks to infer the vehicle state $x_{0}(t), x_{1}(t), x_{2}(t), \cdots, x_{N}(t)$. Under the deterministic quantization, the attacker only observes $\mathcal{Q}_{d}(x_{0}(t)), \mathcal{Q}_{d}(x_{1}(t)), \mathcal{Q}_{d}(x_{2}(t)), \cdots, \mathcal{Q}_{d}(x_{N}(t))$. 

Define the following two signals:
\[
\begin{aligned}
\chi &= \begin{bmatrix}
	\chi_{1}, \chi_{2}, \cdots, \chi_{3(N+1)}
\end{bmatrix}^{\top} 
= \begin{bmatrix}
x_{0}^{\top}, x_{1}^{\top}, \cdots, x_{N}^{\top}
\end{bmatrix}^{\top},
\\
\bar{\chi} &= \begin{bmatrix}
	\bar{\chi}_{1}, \bar{\chi}_{2}, \cdots, \bar{\chi}_{3(N+1)}
\end{bmatrix}^{\top} 
\\ 
&= \begin{bmatrix}
	\mathcal{Q}_{d}(x_{0})^{\top}, \mathcal{Q}_{d}(x_{1})^{\top}, \cdots, \mathcal{Q}_{d}(x_{N})^{\top}
\end{bmatrix}^{\top}.
\end{aligned}
\]
Then, we need to show that $\chi(t)$ cannot be identified from $\bar{\chi}(t)$. According to \eqref{equ:det_quantizer}, we use  
\[
\chi \xRightarrow{\mathcal{Q}_{d}(\cdot), \; 
\Delta} \bar{\chi},
\] 
to denote the transformation from $\chi(t)$ to $\bar{\chi}(t)$ via the deterministic quantizer $\mathcal{Q}_{d}(\cdot)$ with step resolution $\Delta$. For any feasible sequence $\bar{\chi}(t)$ received by the eavesdropper, the set $\Omega(\bar{\chi}(t))$ is defined as
\begin{align*}
	&\Omega(\bar{\chi}) = \lbrace \chi: \exists \left( \mathcal{Q}_{d}(\cdot), \; \Delta \right) \text{s.t.} \; \chi \xRightarrow{\mathcal{Q}_{d}(\cdot), \; 
		\Delta} \bar{\chi} \rbrace.
\end{align*}
Essentially, the set $\Omega(\bar{\chi}(t))$ includes all possible values of $\chi(t)$ that can be transformed into $\bar{\chi}(t)$ with corresponding deterministic quantization scheme~\eqref{equ:det_quantizer}. 

\begin{definition} [$\infty$-Diversity] \label{def_privacy}
	The actual state $\chi(t)$ of the platoon system is said to be privacy-preserving if the cardinality of the set $\Omega(\bar{\chi}(t))$ is infinite for any feasible observation $\bar{\chi}(t)$.  
\end{definition}

The $\infty$-Diversity privacy definition requires that under the deterministic quantizer $\mathcal{Q}_{d}(\cdot)$ and step resolution $\Delta$, there are infinite sets of $\chi(t)$ that can generate the same $\bar{\chi}(t)$ received by the eavesdropper. As a result, it is impossible for the eavesdropper to only use $\bar{\chi}(t)$ to infer the actual state information. 

\begin{remark}
	Definition~\ref{def_privacy} extends the classical $l$-diversity privacy concept~\cite{machanavajjhala2007ACM,Zhang2025AUTO}, which is commonly used in formal analysis of attribute privacy in tabular datasets. In essence, $l$-diversity requires that the privacy-sensitive attributes should have at least $l$ different possible values, with a larger $l$ implying a higher level of indistinguishability.     
\end{remark}

We next show that the deterministic quantization can protect the privacy of the vehicle fleet based on Definition~\ref{def_privacy}.
\begin{theorem} \label{theorem_privacy}
	Under the deterministic quantization mechanism~\eqref{equ:det_quantizer}, the state information $\chi(t)$ is $\infty$-Diversity with respect to any observed $\bar{\chi}(t)$, that is, the eavesdropper cannot infer the actual state information $\chi(t)$ only based on $\bar{\chi}(t)$.
\end{theorem}

\begin{IEEEproof}
	According to Definition~\ref{def_privacy}, we prove Theorem~\ref{theorem_privacy} by showing that, under the deterministic quantizer, the cardinality of the set $\Omega(\bar{\chi}(t))$ is infinite. Specifically, given the quantized signal $\bar{\chi}(t)$ accessible to the attacker, any signal $\chi(t)$ can be mapped into $\bar{\chi}(t)$ through the deterministic quantizer if it satisfies
	\[
	-\frac{\Delta}{2} \le \chi_{\ell} - \bar{\chi}_{\ell} < \frac{\Delta}{2}, \ell= 1,2, \cdots, 3(N+1).
	\]
	Since there are infinitely many $\chi(t)$ that meet this condition, the attacker could receive the same quantized information $\bar{\chi}(t)$ from multiple possible $\chi(t)$. Therefore, the cardinality of the set $\Omega(\bar{\chi}(t))$ is infinite.
\end{IEEEproof}

\begin{remark}
	If the eavesdropper only has access to $\bar{\chi}(t)$, deterministic quantization can offer strong privacy protection by preventing exact inference of the true information $\chi(t)$. However, it is important to note that the $\infty$-Diversity privacy notion is not resilient to auxiliary knowledge. Specifically, if the eavesdropper possesses additional information about the vehicle system and the distributed controller, it may be possible to infer the underlying information even under deterministic quantization. In Section~\ref{sec_perfEva}, we will demonstrate through a simulation case that deterministic quantization lacks robustness when the eavesdropper has access to such auxiliary information. 
\end{remark}

\section{Probabilistic Quantization} \label{sec_probabilistic}
This section presents the distributed platoon control framework under probabilistic quantization, analyzing its stability and privacy-preserving characteristics.

\subsection{Probabilistic Quantizer for Platoon Control}
Instead of directly sharing the actual data with its neighbors, each vehicle uses the probabilistic quantizer to protect the privacy-sensitive information. Specifically, for a vector $z=\begin{bmatrix}
	z_{1}, z_{2}, \cdots, z_{m}
\end{bmatrix}^{\top}\in \mathbb{R}^{m}$, the probabilistic quantizer is given by $\mathcal{Q}_{p}(z) = \begin{bmatrix}
	\mathcal{Q}_{p}(z_{1}), \mathcal{Q}_{p}(z_{2}), \cdots, \mathcal{Q}_{p}(z_{m})
\end{bmatrix}^{\top}$, and $\mathcal{Q}_{p}(z_{\ell})$ ($\ell=1,2, \cdots, m$) is defined as
\begin{equation} \label{equ:pro_quantizer}
	\begin{aligned}
		\mathcal{Q}_{p}(z_{\ell}) &= \begin{cases}
			n\Delta, & \text{with probability}\; \frac{(n+1)\Delta - z_{\ell}}{\Delta}, \\
			(n+1)\Delta, & \text{with probability}\; \frac{z_{\ell}-n\Delta}{\Delta}, 
		\end{cases}
		\\
		& \qquad \qquad \qquad \quad \,\, z_{\ell}\in \left(n\Delta, (n+1)\Delta\right], n\in \mathbb{Z}, 
	\end{aligned}
\end{equation}
where $\Delta>0$ is the quantization step. It follows from~\eqref{equ:pro_quantizer} that $| \mathcal{Q}_{p}(z_{\ell}) - z_{\ell} | \le \Delta$, and some other properties of the probabilistic quantizer are stated in the following lemma.

\begin{lemma}[\cite{Xiao2005TIT}] \label{lemma:prob}
	The probabilistic quantizer~\eqref{equ:pro_quantizer} ensures that $\forall z_{\ell}\in \mathbb{R}$,
	\[
	\begin{aligned}
		\mathbb{E}[\mathcal{Q}_{p}(z_{\ell}) - z_{\ell}] = 0, \quad \mathbb{E}[(\mathcal{Q}_{p}(z_{\ell}) - z_{\ell})^{2}] \le \frac{\Delta^{2}}{4}.
	\end{aligned}
	\]
\end{lemma}

Unlike the deterministic quantizer, the probabilistic quantizer incorporates randomness into the quantization process. For a given input, it selects an output level based on a probability distribution, ensuring that the expected value of the quantized output matches the original input. This unbiasedness property is especially advantageous in distributed control/optimization and machine learning applications, where quantization noise can be mitigated over time or across multiple agents.

The distributed controller under the probabilistic quantization is updated to
\begin{equation} \label{equ:DC_probabilistic}
	\begin{aligned}
		u_{i} = & K\left( \sum_{j=1}^{N} m_{ij}\left( (\mathcal{Q}_{p}(x_{j})+d_{j}) - (\mathcal{Q}_{p}(x_{i})+d_{i}) \right) \right.
		\\
		& \qquad \qquad \qquad \left. + s_{i}\left( \mathcal{Q}_{p}(x_{0}) - (\mathcal{Q}_{p}(x_{i})+d_{i}) \right) \right).
	\end{aligned}
\end{equation}
Let the quantization errors $e_{p0}(t)$ and $e_{pi}(t)$ be defined as
\begin{equation} \label{equ:e_pro}
	\begin{aligned}
		e_{p0} &= \mathcal{Q}_{p}(x_{0})-x_{0}, 
		\\
		e_{pi} &= \mathcal{Q}_{p}(x_{i})-x_{i},\; i=1, 2, \cdots, N.
	\end{aligned}
\end{equation}
Following similar arguments as in Section~\ref{subsec_det_quantizer}, the closed-loop dynamics of the platoon system can be formulated as follows:
\begin{equation} \label{equ:dot_varepsilon_prob}
	\begin{aligned}
		\dot{\varepsilon} &= \left( I_{N}\otimes A - (L+S)\otimes BK \right)\varepsilon 
		\\
		&\qquad \qquad - \left((L+S)\otimes BK \right)\left( e_{p}-1_{N}\otimes e_{p0} \right)
		\\
		&= A_{\varepsilon} \varepsilon  - B_{\varepsilon} \left( e_{p}-1_{N}\otimes e_{p0} \right),
	\end{aligned}
\end{equation}
where $e_{p}(t) = \begin{bmatrix}
	e_{p1}^{\top}(t), e_{p2}^{\top}(t), \cdots, e_{pN}^{\top}(t)
\end{bmatrix}^{\top}$ and $A_{\varepsilon}$, $B_{\varepsilon}$ are defined as
\begin{equation} \label{equ:A_varepsilon}
\begin{aligned}
	A_{\varepsilon} &= I_{N}\otimes A - (L+S)\otimes BK,
    \\
    B_{\varepsilon} &= (L+S)\otimes BK.
\end{aligned}
\end{equation}

\begin{lemma} \label{lemma:prob2}
Let $\bar{e}_{p}(t) = e_{p}(t)-1_{N}\otimes e_{p0}(t)\in \mathbb{R}^{3N}$, then it holds that
\begin{equation} \label{equ:bar_e_varepsilon}
	\begin{aligned}
		\mathbb{E}[\bar{e}_{p}] = 0, \quad \mathbb{E}[\bar{e}_{p}\bar{e}_{p}^{\top}] \le \frac{\Delta^{2}}{4}(N+1)I_{3N}.
	\end{aligned}
\end{equation}
\end{lemma}
\begin{IEEEproof}
	Since the elements of $e_{p}(t)$ and $e_{p0}(t)$ are independent, it can be obtained from Lemma~\ref{lemma:prob} that 
	\begin{equation} \label{equ:E_epro}
	\begin{aligned}
		&\mathbb{E}[e_{p}] = 0, \quad &\mathbb{E}[e_{p}e_{p}^{\top}] \le \frac{\Delta^{2}}{4}I_{3N},
		\\
		&\mathbb{E}[e_{p0}] = 0, \quad &\mathbb{E}[e_{p0}e_{p0}^{\top}] \le \frac{\Delta^{2}}{4}I_{3}.
	\end{aligned}
	\end{equation} 
    Based on~\eqref{equ:E_epro} and $\bar{e}_{p}(t) = e_{p}(t)-1_{N}\otimes e_{p0}(t)$, we have
    \begin{equation}
    \begin{aligned}
    	\mathbb{E}[\bar{e}_{p}] = \mathbb{E}[e_{p}] - \mathbb{E}[1_{N}\otimes e_{p0}] = \mathbb{E}[e_{p}] - 1_{N}\otimes \mathbb{E}[e_{p0}] = 0,
    \end{aligned}
    \end{equation}
    and 
    \begin{equation} \label{equ:E_bar_ep}
    \begin{aligned}
    	\mathbb{E}[\bar{e}_{p}\bar{e}_{p}^{\top}] &= \mathbb{E}[e_{p}e_{p}^{\top}] + \mathbb{E}[(1_{N}\otimes e_{p0})(1_{N}\otimes e_{p0})^{\top}]
    	\\
    	&= \mathbb{E}[e_{p}e_{p}^{\top}] + \mathbb{E}[1_{N}1_{N}^{\top} \otimes e_{p0}e_{p0}^{\top}]
    	\\
    	&= \mathbb{E}[e_{p}e_{p}^{\top}] + 1_{N}1_{N}^{\top} \otimes \mathbb{E}[e_{p0}e_{p0}^{\top}]
    	\\
    	&\le \frac{\Delta^{2}}{4}I_{3N} + 1_{N}1_{N}^{\top} \otimes \frac{\Delta^{2}}{4}I_{3}.
    \end{aligned}
    \end{equation}
    Note that the largest eigenvalue of $1_{N}1_{N}^{\top}$ is $N$, and thus we have $1_{N}1_{N}^{\top} \otimes \frac{\Delta^{2}}{4}I_{3}\le \frac{\Delta^{2}}{4}NI_{3N}$. Based on this inequality and~\eqref{equ:E_bar_ep}, it follows that $\mathbb{E}[\bar{e}_{p}\bar{e}_{p}^{\top}] \le \frac{\Delta^{2}}{4}(N+1)I_{3N}$.
\end{IEEEproof}

\begin{theorem} \label{theorem_prob_cotrol}
	The distributed platoon controller~\eqref{equ:DC_probabilistic} with probabilistic quantization ensures that 
	
	$\left. 1\right)$ $\lim_{t\rightarrow \infty} \mathbb{E}[\varepsilon(t)] = 0$, i.e., the expectation of the collective tracking error $\varepsilon(t)$ converges asymptotically to zero;
	
	
	
	$\left. 2\right)$ $\lim_{t\rightarrow \infty} \mathbb{E}[\varepsilon^{\top}(t)\varepsilon(t)] \le \frac{\Delta^{2}}{4}(N+1)\mathrm{trace}(W)$, where 
	\begin{equation} \label{equ:W}
		W = \int_{0}^{\infty} e^{A_{\varepsilon}\tau}B_{\varepsilon} B_{\varepsilon}^{\top} e^{A^{\top}_{\varepsilon}\tau} d\tau.
	\end{equation}
\end{theorem}

\begin{IEEEproof}
	To prove statement $\left. 1\right)$, we first show that $A_{\varepsilon}$ is Hurwitz. According to~\eqref{equ:A_varepsilon} and the matrix decomposition $L+S = U\Lambda U^{-1}$ in~\eqref{equ:Lambda}, we have
	\begin{equation} \label{equ:A_varepsilon2}
	\begin{aligned}
		A_{\varepsilon} &= I_{N}\otimes A - (L+S)\otimes BK
		\\
		&= I_{N}\otimes A - (U\Lambda U^{-1})\otimes BK
		\\
		&= (U\otimes I_{3}) \left( I_{N}\otimes A - \Lambda \otimes BK \right) (U^{-1}\otimes I_{3}).
	\end{aligned}
	\end{equation}
    The inequality condition in~\eqref{equ:cond1} can be rewritten as 
    \begin{equation} \label{equ:cond3}
    \begin{aligned}
    	&\quad\; I_{N}\otimes (PA+A^{\top}P)-(\Lambda+\Lambda^{\top})\otimes PBB^{\top}P 
    	\\
    	& = (I_{N}\otimes P) \left( I_{N}\otimes A - \Lambda \otimes BK \right) 
    	\\
    	&\quad + \left( I_{N}\otimes A - \Lambda \otimes BK \right)^{\top} (I_{N}\otimes P)
    	\le -\gamma I_{3N}.
    \end{aligned}
    \end{equation} 
    Since $P$ is positive definite, it can be concluded from~\eqref{equ:cond3} that $I_{N}\otimes A - \Lambda \otimes BK$ is Hurwitz. \eqref{equ:A_varepsilon2} indicates that $A_{\varepsilon}$ and $I_{N}\otimes A - \Lambda \otimes BK$ are similar, and thus $A_{\varepsilon}$ is Hurwitz. In addition, the solution to~\eqref{equ:dot_varepsilon_prob} is
    \begin{equation} \label{equ:solution}
    	\varepsilon(t) = e^{A_{\varepsilon}t}\varepsilon(0) - \int_{0}^{t} e^{A_{\varepsilon}(t-\tau)}B_{\varepsilon}\bar{e}_{p}(\tau)d\tau.
    \end{equation} 
    Taking the expectation of~\eqref{equ:solution} and using Lemma~\ref{lemma:prob2}, we have
    \begin{equation} \label{equ:solution_prob}
    	\mathbb{E}[\varepsilon(t)] = e^{A_{\varepsilon}t}\varepsilon(0) - \int_{0}^{t} e^{A_{\varepsilon}(t-\tau)}B_{\varepsilon} \mathbb{E}[\bar{e}_{p}(\tau)]d\tau = e^{A_{\varepsilon}t}\varepsilon(0).
    \end{equation} 
     Since $A_{\varepsilon}$ is Hurwitz, $\lim_{t\rightarrow \infty} \mathbb{E}[\varepsilon(t)] =\lim_{t\rightarrow \infty} e^{A_{\varepsilon}t}\varepsilon(0) = 0$.
     
     We now prove the second statement. The quantify $\mathbb{E}[\varepsilon^{\top}(t)\varepsilon(t)]$ is the trace of the covariance matrix of $\varepsilon(t)$, i.e.,
     \begin{equation} \label{equ:trace}
     	\mathbb{E}[\varepsilon^{\top}(t)\varepsilon(t)] = \mathrm{trace}( \mathbb{E}[\varepsilon(t)\varepsilon^{\top}(t)]). 
     \end{equation}
     From~\eqref{equ:solution} and $\mathbb{E}[\bar{e}_{p}(t)] = 0$, it follows that
     \begin{equation} \label{equ:cova}
     \begin{aligned}
     	&\mathbb{E}[\varepsilon(t)\varepsilon^{\top}(t)] = e^{A_{\varepsilon}t}\varepsilon(0) \varepsilon^{\top}(0)e^{A^{\top}_{\varepsilon}t}  
     	\\
     	&+ \int_{0}^{t} \int_{0}^{t} e^{A_{\varepsilon}(t-\tau_{1})}B_{\varepsilon} \mathbb{E}[\bar{e}_{p}(\tau_{1}) \bar{e}^{\top}_{p}(\tau_{2})] B_{\varepsilon}^{\top} e^{A^{\top}_{\varepsilon}(t-\tau_{2})} d\tau_{1}d\tau_{2}.
     \end{aligned}
     \end{equation}
     Since $\bar{e}_{p}(t)$ is uncorrelated in time, i.e., $\mathbb{E}[\bar{e}_{p}(\tau_{1}) \bar{e}^{\top}_{p}(\tau_{2})]=0$ for $\tau_{1}\neq \tau_{2}$, \eqref{equ:cova} can be simplified to
     \begin{equation}
     \begin{aligned}
     	\mathbb{E}[\varepsilon(t)\varepsilon^{\top}(t)] &= e^{A_{\varepsilon}t}\varepsilon(0) \varepsilon^{\top}(0)e^{A^{\top}_{\varepsilon}t}  
     	\\
     	&+ \int_{0}^{t} e^{A_{\varepsilon}(t-\tau)}B_{\varepsilon} \mathbb{E}[\bar{e}_{p}(\tau) \bar{e}^{\top}_{p}(\tau)] B_{\varepsilon}^{\top} e^{A^{\top}_{\varepsilon}(t-\tau)} d\tau.
     \end{aligned}
     \end{equation}
     From~\eqref{equ:bar_e_varepsilon}, it follows that
     \begin{equation}
     \begin{aligned}
     	\mathbb{E}[\varepsilon(t)\varepsilon^{\top}(t)] &\le e^{A_{\varepsilon}t}\varepsilon(0) \varepsilon^{\top}(0)e^{A^{\top}_{\varepsilon}t}  
     	\\
     	&+ \frac{\Delta^{2}}{4}(N+1) \int_{0}^{t} e^{A_{\varepsilon}(t-\tau)}B_{\varepsilon} B_{\varepsilon}^{\top} e^{A^{\top}_{\varepsilon}(t-\tau)} d\tau.
     \end{aligned}
     \end{equation}
     Since $A_{\varepsilon}$ is Hurwitz, as $t \rightarrow \infty$, the first term vanishes, and then we have 
     \begin{equation} \label{equ:cova2}
     	\lim_{t\rightarrow \infty} \mathbb{E}[\varepsilon(t)\varepsilon^{\top}(t)] \le \frac{\Delta^{2}}{4}(N+1) W,
     \end{equation}
     where $W$ is defined in~\eqref{equ:W} and it is the solution to the Lyapunov equation $A_{\varepsilon}W + WA_{\varepsilon}^{\top} + B_{\varepsilon}B_{\varepsilon}^{\top} = 0$. Based on~\eqref{equ:trace} and~\eqref{equ:cova2}, it can be concluded that $\lim_{t\rightarrow \infty} \mathbb{E}[\varepsilon^{\top}(t)\varepsilon(t)] \le \frac{\Delta^{2}}{4}(N+1)\mathrm{trace}(W)$, which completes the proof.
\end{IEEEproof}

\subsection{Differential Privacy} \label{subsec_DP}
In this subsection, differential privacy is employed to characterize and quantify the privacy guarantees provided by the probabilistic quantizer~\eqref{equ:pro_quantizer}. In particular, $(\epsilon, \delta)$-differential privacy~\cite{Dwork2014FTTCS,cortes2016CDC} offers a probabilistic framework for evaluating the privacy of mechanisms. Some key definitions are provided below. 

\begin{definition}[$\zeta$-Adjacency] \label{definition1}
	Given $\zeta>0$, two state sequences $\chi \in \mathbb{R}^{3(N+1)}$ and $\chi^{\prime} \in \mathbb{R}^{3(N+1)}$ are said to be $\zeta$-adjacent if $\|\chi - \chi^{\prime} \|_{1}\le \zeta $. The set of all such $\zeta$-adjacent pairs is denoted by $\mathrm{Adj}_{1}^{\zeta}$.
\end{definition}


\begin{definition}[$(\epsilon, \delta)$-Differential Privacy] \label{Def:DP} 
	Given $\epsilon$, $\delta\ge 0$, a random mechanism $\mathcal{M}$ is said to satisfy $(\epsilon, \delta)$-differential privacy if, for any $\mathcal{S}\subseteq \mathrm{range}(\mathcal{M})$ and for any $(\chi, \chi^{\prime}) \in \mathrm{Adj}_{1}^{\zeta}$, the following holds:
	\begin{equation}\label{equ:DP}
		\mathbb{P}(\mathcal{M}(\chi) \in  \mathcal{S}) \leq e^\epsilon \mathbb{P}\left( \mathcal{M} \left(\chi^{\prime}\right) \in  \mathcal{S}\right) + \delta.
	\end{equation}
\end{definition}

Definition~\ref{Def:DP} implies that for two $\zeta$-adjacent state sequences $\chi$ and $\chi^{\prime}$, a mechanism $\mathcal{M}(\cdot)$ is differentially private if it ensures that the outputs of the two sequences are different in probabilities by at most $\epsilon$ and $\delta$ specified on the right hand side of~\eqref{equ:DP}. The parameters $\epsilon$ and $\delta$ quantify how distinguishable the outputs are for adjacent inputs.
A smaller $\epsilon$ or $\delta$ indicates that the mechanism makes adjacent sequences less distinguishable, thereby providing stronger privacy guarantees.

\begin{theorem} \label{theorem_DP}
	Given $0<\zeta < \Delta$, the probabilistic quantization mechanism described in \eqref{equ:pro_quantizer} can achieve $(0, \frac{\zeta}{\Delta})$-differential privacy for any $(\chi, \chi^{\prime}) \in \mathrm{Adj}_{1}^{\zeta}$.
\end{theorem}
\begin{IEEEproof}
    Since the quantization of each element is independent of the others—that is, the quantization errors across different elements are mutually independent—we can analyze the privacy of each component of $\chi(t)$ separately. According to Definition~\ref{Def:DP}, to establish that the mechanism achieves $(0, \frac{\zeta}{\Delta})$-differential privacy, it suffices to show that
    $\left| \mathbb{P}(\mathcal{Q}_{p}(\chi_{\ell}) \in  \mathcal{S}| \chi) - \mathbb{P}(\mathcal{Q}_{p}(\chi_{\ell}^{\prime}) \in  \mathcal{S}| \chi^{\prime}) \right| \le \frac{\zeta}{\Delta}$ for all $\chi$, $\chi^{\prime}$ such that $\|\chi - \chi^{\prime} \|_{1}\le \zeta $. The condition $\|\chi - \chi^{\prime} \|_{1}\le \zeta $ implies that $| \chi_{\ell} - \chi^{\prime}_{\ell} | \le \zeta < \Delta$. To proceed, we consider two cases in the derivation: $\left. 1 \right)$ $\chi_{\ell}$, $\chi_{\ell}^{\prime}\in \left(n\Delta, (n+1)\Delta\right]$; $\left. 2 \right)$ $\chi_{\ell}\in \left(n\Delta, (n+1)\Delta\right]$ and $\chi_{\ell}^{\prime}\in \left((n+1)\Delta, (n+2)\Delta\right]$.
	
	\textbf{Case 1}: When $\chi_{\ell}$, $\chi_{\ell}^{\prime}\in \left(n\Delta, (n+1)\Delta\right]$, we have 
	\[
	\mathcal{S} \subseteq \left\lbrace n\Delta, (n+1)\Delta \right\rbrace. 
	\]
	\begin{itemize}
		\item For $\mathcal{S}=\left\lbrace n\Delta \right\rbrace$, it follows from~\eqref{equ:pro_quantizer} that
		\[
		\begin{aligned}
		&\sup_{\|\chi-\chi^{\prime}\|_{1} \le \zeta} \left| \mathbb{P}(\mathcal{Q}_{p}(\chi_{\ell}) =n\Delta| \chi) - \mathbb{P}(\mathcal{Q}_{p}(\chi_{\ell}^{\prime}) =n\Delta| \chi^{\prime}) \right|
		\\
		&= \sup_{\|\chi-\chi^{\prime}\|_{1} \le \zeta} \left| \frac{(n+1)\Delta - \chi_{\ell}}{\Delta} - \frac{(n+1)\Delta - \chi_{\ell}^{\prime}}{\Delta} \right|
		\\
		&= \sup_{\|\chi-\chi^{\prime}\|_{1} \le \zeta} \left| \frac{ \chi_{\ell}^{\prime}-\chi_{\ell}}{\Delta} \right|
		\le \frac{\|\chi-\chi^{\prime}\|_{1}}{\Delta} \le \frac{\zeta}{\Delta}.
		\end{aligned}
		\]
		\\
		\item For $\mathcal{S}=\left\lbrace (n+1)\Delta \right\rbrace$, we have
		\[
		\begin{aligned}
			&\sup_{\|\chi-\chi^{\prime}\|_{1} \le \zeta} \left| \mathbb{P}(\mathcal{Q}_{p}(\chi_{\ell}) =(n+1)\Delta| \chi) \right.
			\\
			&\left. \qquad \qquad \qquad - \mathbb{P}(\mathcal{Q}_{p}(\chi_{\ell}^{\prime}) =(n+1)\Delta| \chi^{\prime}) \right|
			\\
			&= \sup_{\|\chi-\chi^{\prime}\|_{1} \le \zeta} \left| \frac{\chi_{\ell}-n\Delta}{\Delta} - \frac{\chi_{\ell}^{\prime}-n\Delta}{\Delta} \right|
			\\
			&= \sup_{\|\chi-\chi^{\prime}\|_{1} \le \zeta} \left| \frac{\chi_{\ell}- \chi_{\ell}^{\prime}}{\Delta} \right|
			\le \frac{\|\chi-\chi^{\prime}\|_{1}}{\Delta} \le \frac{\zeta}{\Delta}.
		\end{aligned}
		\]
		\\
		\item For $\mathcal{S}=\emptyset$ or $\mathcal{S}=\left\lbrace n\Delta, (n+1)\Delta \right\rbrace$, it holds that
		\[
		\mathbb{P}(\mathcal{Q}_{p}(\chi_{\ell}) \in  \mathcal{S}| \chi) - \mathbb{P}(\mathcal{Q}_{p}(\chi_{\ell}^{\prime}) \in  \mathcal{S}| \chi^{\prime}) = 0 \le \frac{\zeta}{\Delta}.
		\]
	\end{itemize}

    \textbf{Case 2}: When $\chi_{\ell}\in \left(n\Delta, (n+1)\Delta\right]$ and $\chi_{\ell}^{\prime}\in \left((n+1)\Delta, (n+2)\Delta\right]$, we have 
    \[
    \mathcal{S} \subseteq \left\lbrace n\Delta, (n+1)\Delta, (n+2)\Delta \right\rbrace. 
    \]
    \begin{itemize}
    	\item For $\mathcal{S}=\left\lbrace n\Delta \right\rbrace$, it follows that
    	\[
    	\begin{aligned}
    		&\sup_{\|\chi-\chi^{\prime}\|_{1} \le \zeta} \left| \mathbb{P}(\mathcal{Q}_{p}(\chi_{\ell}) =n\Delta| \chi) - \mathbb{P}(\mathcal{Q}_{p}(\chi_{\ell}^{\prime}) =n\Delta| \chi^{\prime}) \right|
    		\\
    		&= \sup_{\|\chi-\chi^{\prime}\|_{1} \le \zeta} \left| \frac{(n+1)\Delta - \chi_{\ell}}{\Delta} - 0 \right|
    		\\
    		&\le \sup_{\|\chi-\chi^{\prime}\|_{1} \le \zeta} \left| \frac{ \chi_{\ell}^{\prime}-\chi_{\ell}}{\Delta} \right|
    		\le \frac{\|\chi-\chi^{\prime}\|_{1}}{\Delta} \le \frac{\zeta}{\Delta},
    	\end{aligned}
    	\]
    	where the first inequality is derived based on $\chi_{\ell}\le (n+1)\Delta < \chi_{\ell}^{\prime}$.
    	\\
    	\item For $\mathcal{S}=\left\lbrace (n+1)\Delta \right\rbrace$, we have
    	\[
    	\begin{aligned}
    		&\sup_{\|\chi-\chi^{\prime}\|_{1} \le \zeta} \left| \mathbb{P}(\mathcal{Q}_{p}(\chi_{\ell}) =(n+1)\Delta| \chi) \right.
    		\\
    		&\left. \qquad \qquad \qquad - \mathbb{P}(\mathcal{Q}_{p}(\chi_{\ell}^{\prime}) =(n+1)\Delta| \chi^{\prime}) \right|
    		\\
    		&= \sup_{\|\chi-\chi^{\prime}\|_{1} \le \zeta} \left| \frac{\chi_{\ell}-n\Delta}{\Delta} - \frac{(n+2)\Delta-\chi_{\ell}^{\prime}}{\Delta} \right|
    		\\
    		&= \sup_{\|\chi-\chi^{\prime}\|_{1} \le \zeta} \left| \frac{ (\chi_{\ell}^{\prime} - \chi_{\ell}) + (2\chi_{\ell} -2(n+1)\Delta)}{\Delta} \right|
    		\\
    		&\le \sup_{\|\chi-\chi^{\prime}\|_{1} \le \zeta} \left| \frac{ \chi_{\ell}^{\prime}-\chi_{\ell}}{\Delta} \right|
    		\le \frac{\|\chi-\chi^{\prime}\|_{1}}{\Delta} \le \frac{\zeta}{\Delta}.
    	\end{aligned}
    	\]
    	\\
    	\item For $\mathcal{S}=\left\lbrace (n+2)\Delta \right\rbrace$, the same result can be obtained by following the similar arguments in the case where $\mathcal{S}=\left\lbrace n\Delta \right\rbrace$.
    	\\
    	\item For $\mathcal{S}=\left\lbrace n\Delta, (n+1)\Delta \right\rbrace$, it holds that
    	\[
    	\begin{aligned}
    		&\sup_{\|\chi-\chi^{\prime}\|_{1} \le \zeta} \left| \mathbb{P}(\mathcal{Q}_{p}(\chi_{\ell}) \in \mathcal{S}| \chi) - \mathbb{P}(\mathcal{Q}_{p}(\chi_{\ell}^{\prime}) \in \mathcal{S}| \chi^{\prime}) \right|
    		\\
    		&= \sup_{\|\chi-\chi^{\prime}\|_{1} \le \zeta} \left| 1 - \frac{(n+2)\Delta-\chi_{\ell}^{\prime}}{\Delta} \right|
    		\\
    		&= \sup_{\|\chi-\chi^{\prime}\|_{1} \le \zeta} \left| \frac{\chi_{\ell}^{\prime}-(n+1)\Delta}{\Delta} \right|
    		\\
    		&\le \sup_{\|\chi-\chi^{\prime}\|_{1} \le \zeta} \left| \frac{ \chi_{\ell}^{\prime}-\chi_{\ell}}{\Delta} \right|
    		\le \frac{\|\chi-\chi^{\prime}\|_{1}}{\Delta} \le \frac{\zeta}{\Delta}.
    	\end{aligned}
    	\]
    	\\
    	\item For $\mathcal{S}=\left\lbrace (n+1)\Delta, (n+2)\Delta \right\rbrace$, we have 
    	\[
    	\begin{aligned}
    		&\sup_{\|\chi-\chi^{\prime}\|_{1} \le \zeta} \left| \mathbb{P}(\mathcal{Q}_{p}(\chi_{\ell}) \in \mathcal{S}| \chi) - \mathbb{P}(\mathcal{Q}_{p}(\chi_{\ell}^{\prime}) \in \mathcal{S}| \chi^{\prime}) \right|
    		\\
    		&= \sup_{\|\chi-\chi^{\prime}\|_{1} \le \zeta} \left| \frac{\chi_{\ell}-n\Delta}{\Delta} - 1 \right| 
    		\\
    		&= \sup_{\|\chi-\chi^{\prime}\|_{1} \le \zeta} \left| \frac{(n+1)\Delta - \chi_{\ell}}{\Delta} \right|
    		\\
    		&\le \sup_{\|\chi-\chi^{\prime}\|_{1} \le \zeta} \left| \frac{ \chi_{\ell}^{\prime}-\chi_{\ell}}{\Delta} \right|
    		\le \frac{\|\chi-\chi^{\prime}\|_{1}}{\Delta} \le \frac{\zeta}{\Delta}.
    	\end{aligned}
    	\]
    	\\
    	\item For $\mathcal{S}=\left\lbrace n\Delta, (n+2)\Delta \right\rbrace$, we have 
    	\[
    	\begin{aligned}
    		&\sup_{\|\chi-\chi^{\prime}\|_{1} \le \zeta} \left| \mathbb{P}(\mathcal{Q}_{p}(\chi_{\ell}) \in \mathcal{S}| \chi) - \mathbb{P}(\mathcal{Q}_{p}(\chi_{\ell}^{\prime}) \in \mathcal{S}| \chi^{\prime}) \right|
    		\\
    		&= \sup_{\|\chi-\chi^{\prime}\|_{1} \le \zeta} \left| \frac{(n+1)\Delta - \chi_{\ell}}{\Delta} - \frac{\chi_{\ell}^{\prime}-(n+1)\Delta}{\Delta}  \right| 
    		\\
    		&= \sup_{\|\chi-\chi^{\prime}\|_{1} \le \zeta} \left| \frac{(2(n+1)\Delta - 2\chi_{\ell}^{\prime}) + (\chi_{\ell}^{\prime}-\chi_{\ell}) }{\Delta} \right|
    		\\
    		&\le \sup_{\|\chi-\chi^{\prime}\|_{1} \le \zeta} \left| \frac{ \chi_{\ell}^{\prime}-\chi_{\ell}}{\Delta} \right|
    		\le \frac{\|\chi-\chi^{\prime}\|_{1}}{\Delta} \le \frac{\zeta}{\Delta}.
    	\end{aligned}
    	\]
    	\\
    	\item For $\mathcal{S}=\emptyset$ or $\mathcal{S}=\left\lbrace n\Delta, (n+1)\Delta, (n+1)\Delta \right\rbrace$, it holds that
    	\[
    	\mathbb{P}(\mathcal{Q}_{p}(\chi_{\ell}) \in  \mathcal{S}| \chi) - \mathbb{P}(\mathcal{Q}_{p}(\chi_{\ell}^{\prime}) \in  \mathcal{S}| \chi^{\prime}) = 0 \le \frac{\zeta}{\Delta}.
    	\]
    \end{itemize}

    Based on the results in Case 1 and Case 2, it can be concluded that $\left| \mathbb{P}(\mathcal{Q}_{p}(\chi_{\ell}) \in  \mathcal{S}| \chi) - \mathbb{P}(\mathcal{Q}_{p}(\chi_{\ell}^{\prime}) \in  \mathcal{S}| \chi^{\prime}) \right| \le \frac{\zeta}{\Delta}$ for any $(\chi, \chi^{\prime})\in \mathrm{Adj}_{1}^{\zeta}$. Therefore, the probabilistic quantizer guarantees $(\epsilon, \delta)$-differential privacy with $\epsilon=0$ and $\delta=\frac{\zeta}{\Delta}$. 
\end{IEEEproof}

\begin{remark}
    The key difference between deterministic and probabilistic quantizers lies in how they handle quantization error and their resulting statistical properties. The deterministic quantizer produces fixed, often biased errors that can accumulate or correlate with the input, potentially degrading system performance or convergence. In contrast, the probabilistic quantizer introduces random, zero-mean errors that are statistically independent of the input in expectation, thereby preserving accuracy in aggregate computations and improving robustness in distributed settings.
\end{remark}

\begin{remark}
	The deterministic quantizer ensures $\infty$-Diversity, which protects privacy by guaranteeing that for any observed $\bar{\chi}(t)$ (i.e., $\mathcal{Q}_d(\chi(t))$), there exist infinitely many possible values of $\chi(t)$ that could result in the same quantized output. This makes it difficult for an attacker to infer the true value of $\chi(t)$ from $\bar{\chi}(t)$. However, $\infty$-Diversity may be vulnerable when an adversary possesses auxiliary information. In contrast, the probabilistic quantizer offers differential privacy, which is a fundamentally stronger and more flexible guarantee. 
    Differential privacy ensures that the output of a mechanism remains approximately the same, whether or not any individual's data is changed.
    It can prevent privacy leakage from a wide range of adversaries, including those with access to auxiliary information.
\end{remark}

\subsection{Trade-off Between Control and Privacy} \label{subsec_trade-off}
In this subsection, we investigate the trade-off between control performance and privacy protection.
Theorem~\ref{theorem_prob_cotrol} shows that $\lim_{t\rightarrow \infty} \mathbb{E}[\varepsilon^{\top}(t)\varepsilon(t)] \le \frac{\Delta^{2}}{4}(N+1)\mathrm{trace}(W)$, indicating that a smaller quantization step $\Delta$ leads to better control performance.
On the other hand, as shown in Theorem~\ref{theorem_DP}, the probabilistic quantizer provides $(0, \delta)$-differential privacy with $\delta=\frac{\zeta}{\Delta}$. Hence, increasing the quantization step $\Delta$ leads to a smaller $\delta$, offering stronger privacy guarantees. 
To balance this trade-off, an optimization problem is formulated. Specifically, since $\lim_{t\rightarrow \infty} \mathbb{E}[\varepsilon^{\top}(t)\varepsilon(t)] \propto \Delta^{2}$ and $\delta \propto \frac{1}{\Delta}$, two objective functions are defined as
\begin{equation} \label{equ:f1_f2}
\begin{aligned}
	f_{1} = \Delta^{2}, \quad f_{2} = \frac{1}{\Delta}, \quad \Delta>0.
\end{aligned}
\end{equation}   
There is no single value of $\Delta$ that minimizes both objective functions simultaneously. Instead, the trade-off can be characterized using the Pareto front~\cite{marler2010SMO}, which consists of all non-dominated solutions. Given~\eqref{equ:f1_f2}, the Pareto front in the objective space is given by $f_{1} = (\frac{1}{f_{2}})^{2}$, $f_{2}>0$. This curve defines the best trade-offs one can achieve between control and privacy: improving one objective inevitably compromises the other.

To choose a specific solution from the Pareto front based on application requirements, a weighted sum optimization problem can be formulated:
\begin{equation} \label{equ:f}
\begin{aligned}
	\min_{\Delta>0} f(\Delta)= w_{1}f_{1} + w_{2}f_{2} = w_{1}\Delta^{2} + w_{2}\frac{1}{\Delta},
\end{aligned}
\end{equation}
where $w_{1}$, $w_{2}>0$ are user-defined weighting factors that reflect the relative importance of control and privacy. Given $w_{1}$ and $w_{2}$, the optimal solution to~\eqref{equ:f} lies on the Pareto front and represents a balanced trade-off between the two competing objectives. 

\begin{figure}[!t]
	\setlength{\abovecaptionskip}{0pt}
	\centering
	\subfigure[] {\label{fig_DetQ_BD}
		\includegraphics[width=0.24\textwidth]{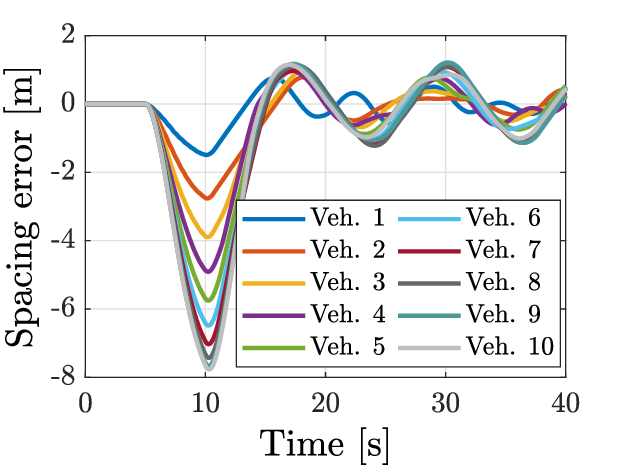}
	}
	\hspace{-0.225 in}
	\subfigure[] {\label{fig_DetQ_BDL}
		\includegraphics[width=0.24\textwidth]{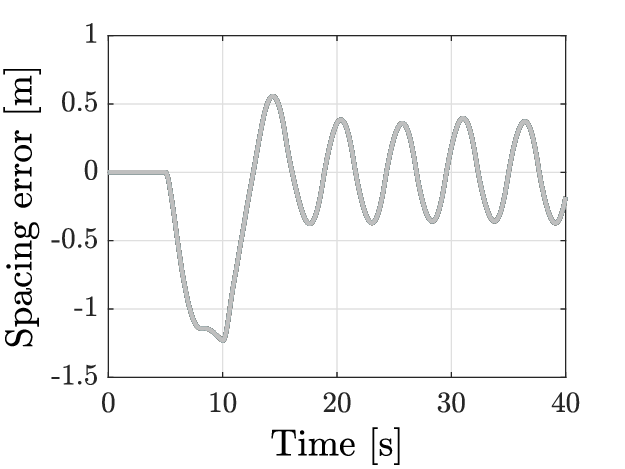}
	}
	\hspace{-0.225 in}
	\subfigure[]{\label{fig_DetQ_PF}
		\includegraphics[width=0.24\textwidth]{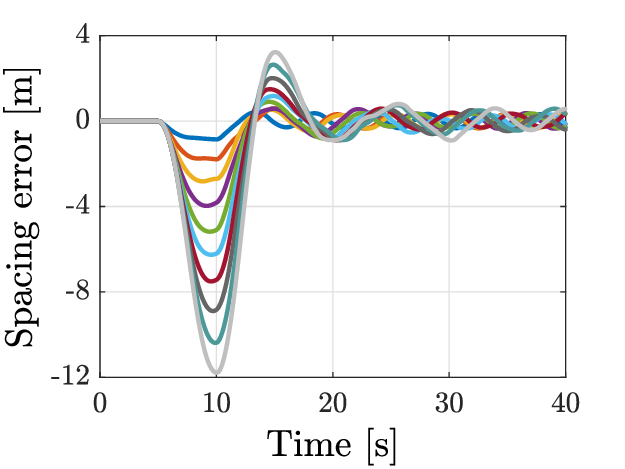}
	}
    \hspace{-0.225 in}
    \subfigure[]{\label{fig_DetQ_PLF}
	    \includegraphics[width=0.24\textwidth]{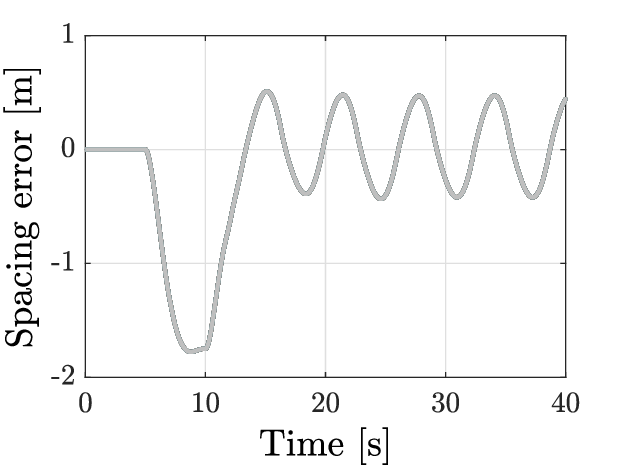}
    }
    \hspace{-0.225 in}
    \subfigure[]{\label{fig_DetQ_TPF}
	    \includegraphics[width=0.24\textwidth]{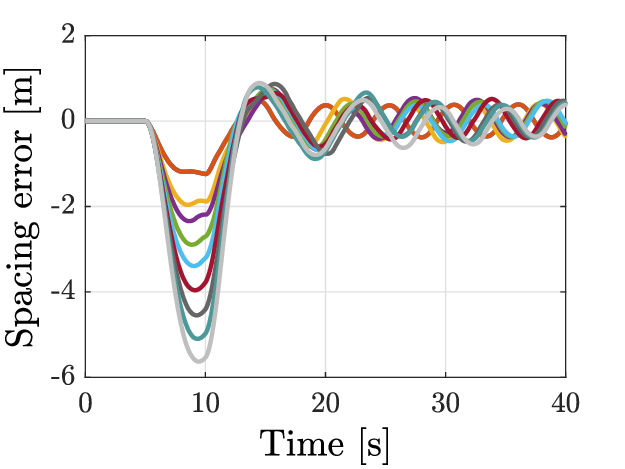}
    }
    \hspace{-0.225 in}
    \subfigure[]{\label{fig_DetQ_TPLF}
	    \includegraphics[width=0.24\textwidth]{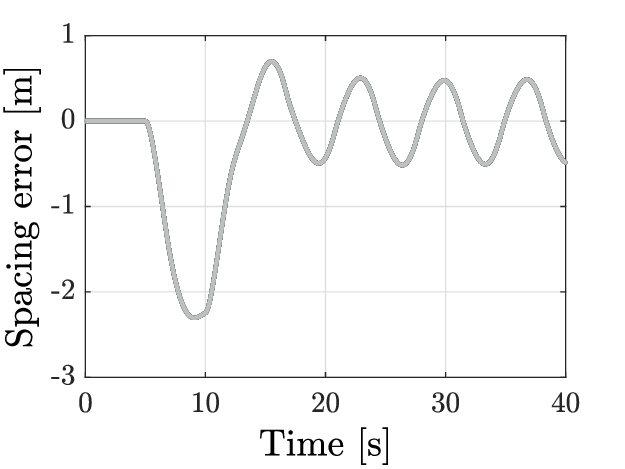}
    }
	\caption{Performance of the distributed platoon controller~\eqref{equ:DC_deterministic} with deterministic quantization ($\Delta=1$): (a) BD, (b) BDL, (c) PF, (d) PLF, (e) TPF, and (f) TPLF.}
	\label{fig_DetQ}
\end{figure}

\begin{figure}[!t]
	\setlength{\abovecaptionskip}{0pt}
	\centering
	\subfigure[] {\label{fig_ProQ_BD}
		\includegraphics[width=0.24\textwidth]{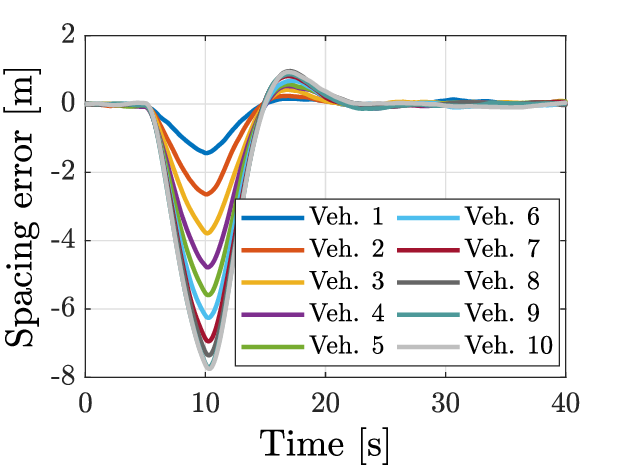}
	}
	\hspace{-0.225 in}
	\subfigure[] {\label{fig_ProQ_BDL}
		\includegraphics[width=0.24\textwidth]{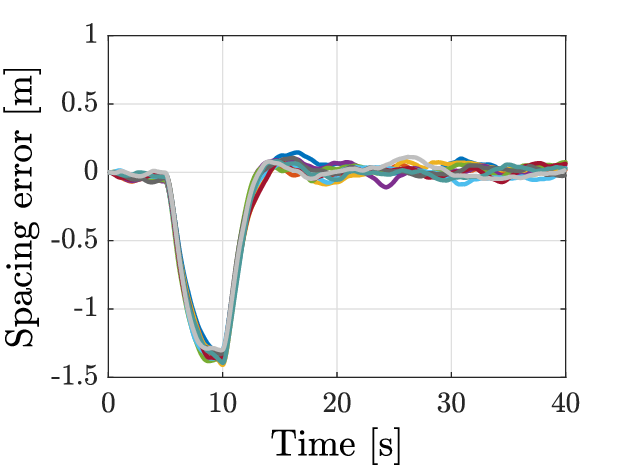}
	}
	\hspace{-0.225 in}
	\subfigure[]{\label{fig_ProQ_PF}
		\includegraphics[width=0.24\textwidth]{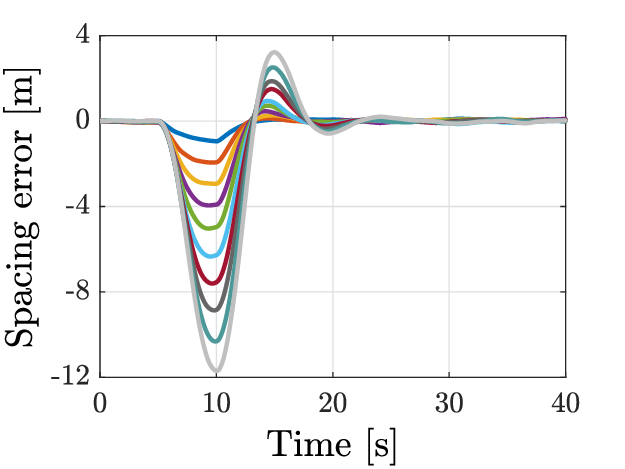}
	}
	\hspace{-0.225 in}
	\subfigure[]{\label{fig_ProQ_PLF}
		\includegraphics[width=0.24\textwidth]{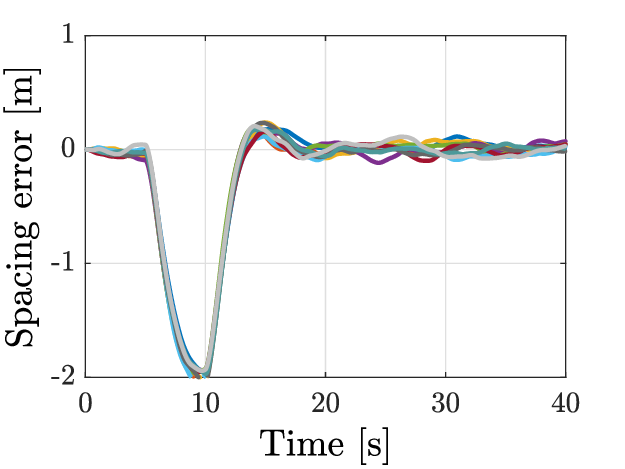}
	}
    \hspace{-0.225 in}
    \subfigure[]{\label{fig_ProQ_TPF}
    	\includegraphics[width=0.24\textwidth]{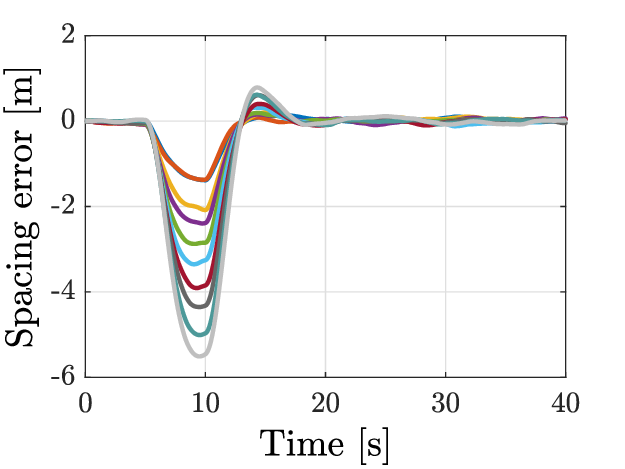}
    }
    \hspace{-0.225 in}
    \subfigure[]{\label{fig_ProQ_TPLF}
    	\includegraphics[width=0.24\textwidth]{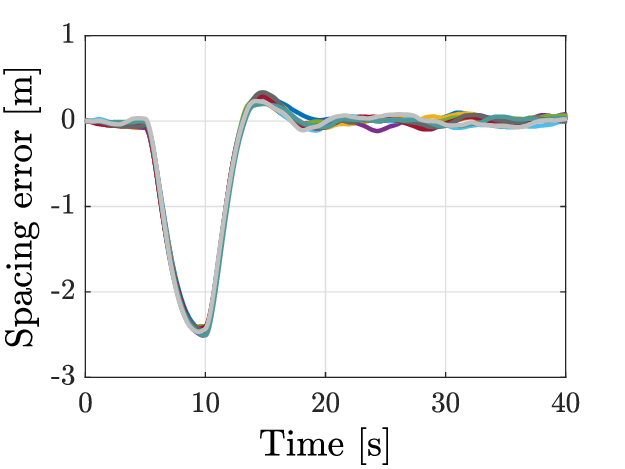}
    }
	\caption{Performance of the distributed platoon controller~\eqref{equ:DC_probabilistic} with probabilistic quantization ($\Delta=1$): (a) BD, (b) BDL, (c) PF, (d) PLF, (e) TPF, and (f) TPLF.}
	\label{fig_ProQ}
\end{figure}

\section{Numerical Simulations} \label{sec_perfEva}
To evaluate the effectiveness of the distributed platoon control strategies under both deterministic and probabilistic quantization, we perform a series of numerical simulations. The scenario involves a homogeneous platoon consisting of 11 identical vehicles—1 lead vehicle and 10 followers—organized according to the communication topologies depicted in Fig.~\ref{fig:SystemSchematic}. The desired inter-vehicle spacing is fixed at $d_{r} = 20$m. In this setup, variations in the lead vehicle's acceleration or deceleration are treated as external disturbances affecting the platoon dynamics. The initial position of the lead vehicle is set to $p_0(0) = 0$, and its velocity profile over time is defined as
\[
v_{0} = \begin{cases}
	20 \;\mathrm{m/s}, & t\le 5 \mathrm{s},
	\\
	20+2t \;\mathrm{m/s}, & 5 \mathrm{s} < t \le 10 \mathrm{s},
	\\
	30 \;\mathrm{m/s}, & t> 10 \mathrm{s}.
\end{cases}
\]
This velocity trajectory introduces a gradual speed increase, simulating a realistic disturbance scenario for assessing control and spacing performance across the platoon.

\begin{figure*}[!t]
	\setlength{\abovecaptionskip}{0pt}
	\centering
	\subfigure[] {\label{fig_DetQ_Delta}
		\includegraphics[width=0.4\textwidth]{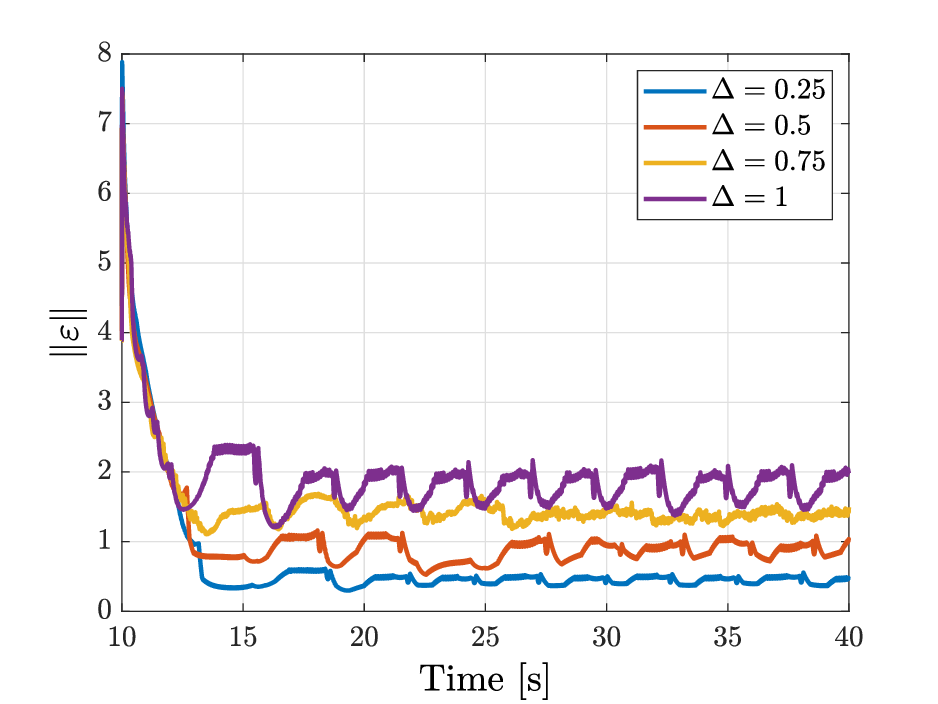}
	}
	\hspace{0 in}
	\subfigure[] {\label{fig_ProQ_Delta}
		\includegraphics[width=0.4\textwidth]{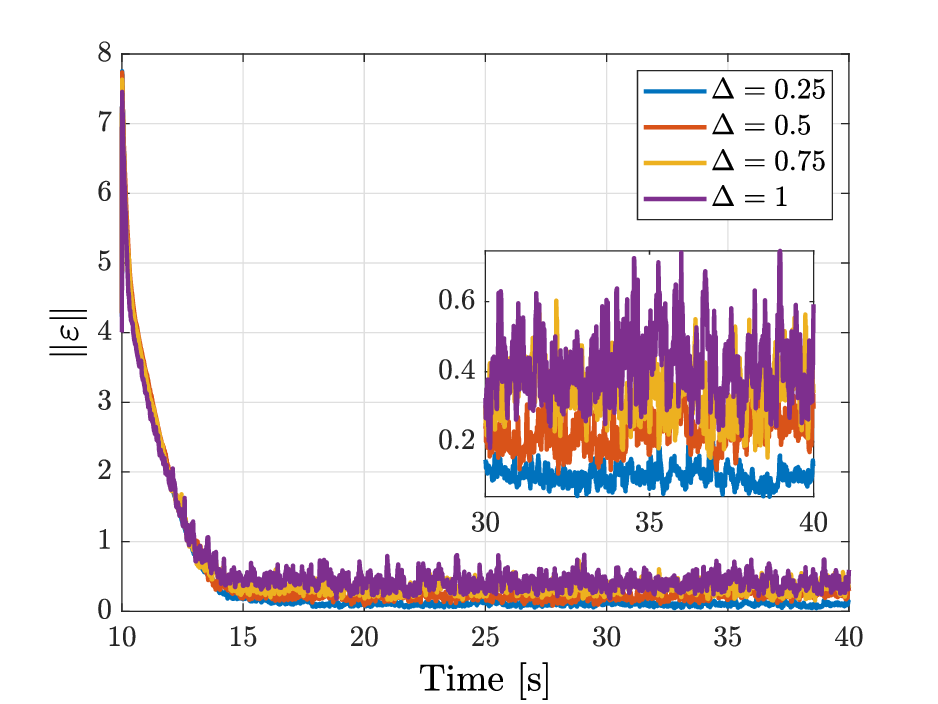}
	}
	\caption{Tracking errors of distributed platoon controllers with BDL topology for $\Delta = 0.25, 0.5, 0.75, 1$: (a) Controller~\eqref{equ:DC_deterministic} with deterministic quantization; (b) Controller~\eqref{equ:DC_probabilistic} with probabilistic quantization.}
	\label{fig_Delta}
\end{figure*}

\begin{figure*}[!t]
	\setlength{\abovecaptionskip}{0pt}
	\centering
	\subfigure[] {\label{fig_DetQ_est}
		\includegraphics[width=0.4\textwidth]{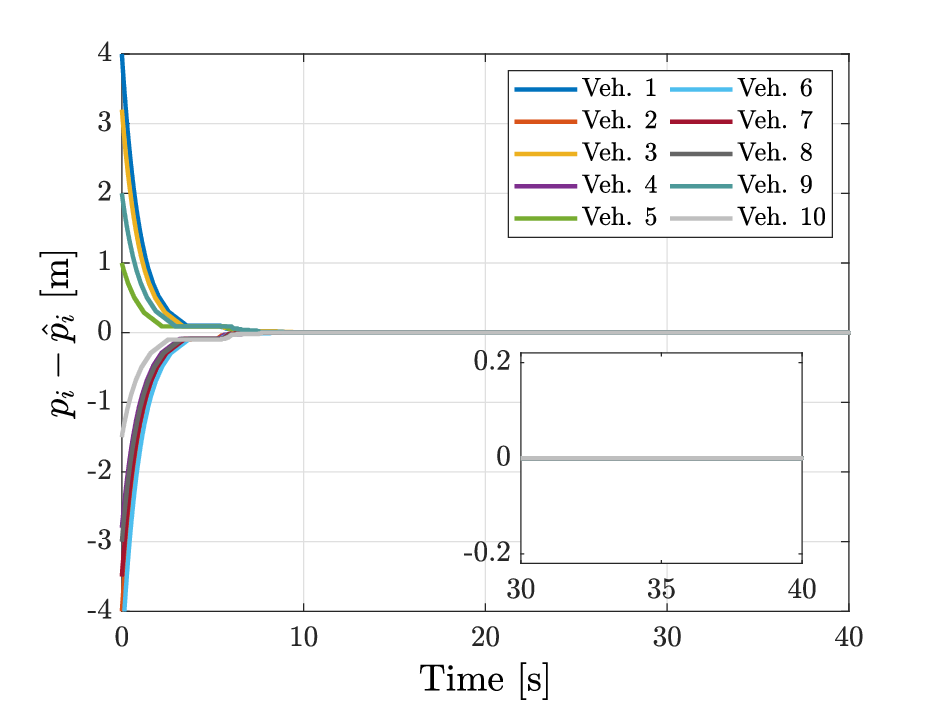}
	}
	\hspace{0 in}
	\subfigure[] {\label{fig_ProQ_est}
		\includegraphics[width=0.4\textwidth]{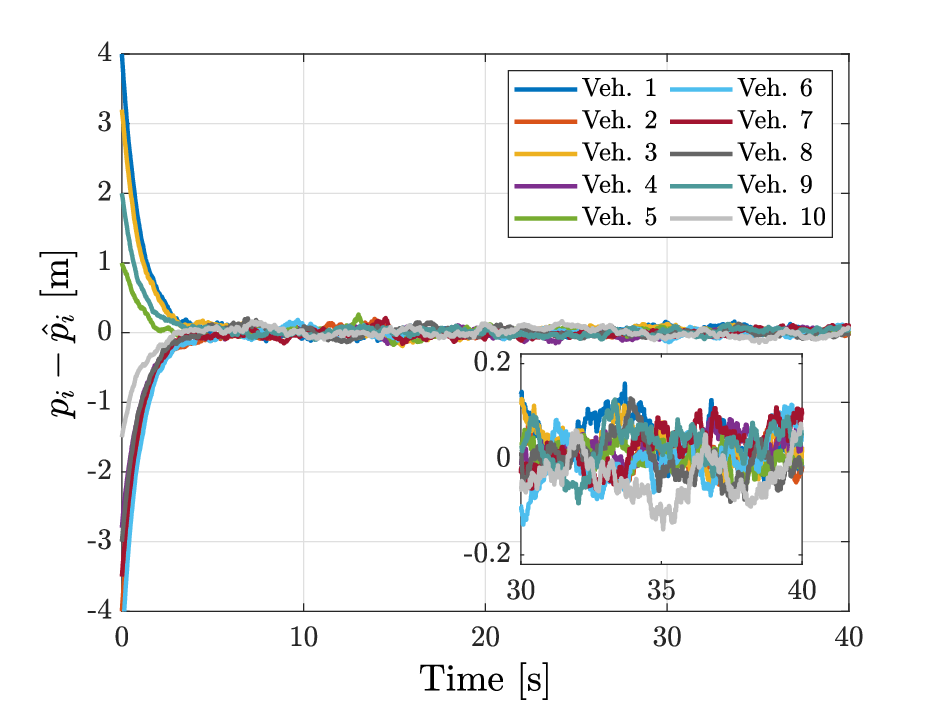}
	}
	\caption{Privacy protection performance of two quantization approaches with the eavesdropper using the estimation scheme in~\eqref{equ:estimator}, under the BD topology: (a) Deterministic quantizer applied to platoon control; (b) Probabilistic quantizer applied to platoon control.}
	\label{fig_estimation}
\end{figure*}

\subsection{Control Performance Validation}
Both distributed platoon controllers under the deterministic quantizer~\eqref{equ:DC_deterministic} and the probabilistic quantizer~\eqref{equ:DC_probabilistic} are tested using the communication topologies illustrated in Figs.~\ref{fig:BD}-\ref{fig:TPLF}. In both cases, the quantization step is set to $\Delta = 1$. The simulation results are presented in~Figs.~\ref{fig_DetQ} and \ref{fig_ProQ}, where the spacing error is defined as $p_{i}(t) + id_{r}-p_{0}(t)$. 
It can be seen that the deterministic quantizer results in spacing errors that oscillate significantly around zero, indicating less stable convergence. In contrast, the probabilistic quantizer effectively suppresses fluctuations and achieves more precise and stable regulation. These results suggest that, compared to its deterministic counterpart, the probabilistic quantizer introduces less disturbance into the system and achieves better control performance.

To further examine how the quantization step affects control accuracy, both controllers are evaluated under the BDL topology using different step sizes: $\Delta = 0.25$, $0.5$, $0.75$, and $1$. The corresponding collective tracking errors across the entire platoon are plotted in Fig.~\ref{fig_Delta}. The results indicate a clear trend: a larger quantization step leads to an increased tracking error for both controllers. Moreover, the probabilistic quantizer consistently outperforms the deterministic one by maintaining lower tracking errors across all tested step sizes.	

\subsection{Privacy Protection Validation}
As discussed in Section~\ref{subsec_diversity}, the deterministic quantizer can preserve privacy when the eavesdropper has access only to the quantized signals transmitted over the communication network. However, this scheme becomes vulnerable when the adversary possesses auxiliary knowledge. In contrast, Section~\ref{subsec_DP} demonstrates that the probabilistic quantizer satisfies differential privacy, ensuring protection even when the eavesdropper has additional background information. To illustrate the contrast in privacy protection between these two quantizers, we introduce an eavesdropping scenario. Specifically, we assume the eavesdropper has full access not only to all quantized signals but also to the system matrices ($A$ and $B$), the communication topology, and the control algorithms. Leveraging this comprehensive information, the eavesdropper employs the following estimator to reconstruct the private state $x_{i}(t)$:
\begin{equation} \label{equ:estimator}
\begin{aligned}
	\dot{\hat{x}}_{i} = A\hat{x}_{i} + Bu_{i} + C(\mathcal{Q}(x_{i})-\mathcal{Q}(\hat{x}_{i})),
\end{aligned}
\end{equation}
where $\hat{x}_{i}(t) = \begin{bmatrix}
	\hat{p}_{i}(t), \hat{u}_{i}(t), \hat{a}_{i}(t)
\end{bmatrix}^{\top}$ is the estimate of $x_{i}(t) = \begin{bmatrix}
p_{i}(t), u_{i}(t), a_{i}(t)
\end{bmatrix}^{\top}$, $C=A+I_{3}$, and $\mathcal{Q}(\cdot)$ represents either the deterministic quantizer $\mathcal{Q}_{d}(\cdot)$ or the probabilistic quantizer $\mathcal{Q}_{p}(\cdot)$ depending on the control implementation. 

To evaluate the privacy-preserving performance of the two schemes, simulations are conducted using the BD topology. The results, shown in Fig.\ref{fig_estimation}, reveal that under the deterministic quantizer, the eavesdropper can successfully reconstruct the target state using estimator\eqref{equ:estimator}. In contrast, the stochastic nature of the probabilistic quantizer prevents accurate inference, rendering the eavesdropper's estimation ineffective. This highlights the advantage of the probabilistic approach in providing stronger privacy guarantees, especially when adversaries possess detailed knowledge of the platoon system.

\begin{figure}[t]
	\centering
	\includegraphics[width=0.4\textwidth]{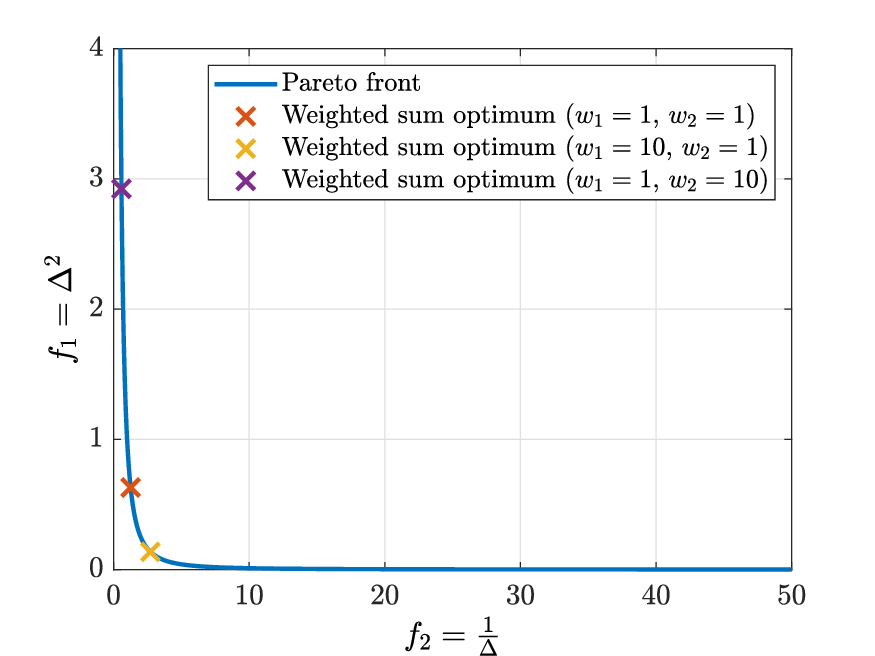}
	\caption{Pareto front illustrating the trade-off between control performance and privacy.}
	\label{fig_pareto}
\end{figure}

We finally illustrate the trade-off between the control performance and privacy discussed in Section~\ref{subsec_trade-off}. In the case of the probabilistic quantizer, increasing the quantization step $\Delta$ enhances privacy guarantees but degrades control accuracy. Fig.~\ref{fig_pareto} presents the Pareto front of the two objective functions defined in~\eqref{equ:f1_f2}. By selecting appropriate weighting factors $w_1$ and $w_2$, one can determine the optimal quantization step $\Delta$ by solving the optimization problem in~\eqref{equ:f}. Fig.~\ref{fig_pareto} also illustrates the resulting solutions corresponding to three different pairs of $(w_1, w_2)$, highlighting the impact of different trade-off preferences.

\section{Conclusion} \label{sec_conclusion}
This paper has studied the stability and privacy-preserving properties of distributed platoon control under both deterministic and probabilistic quantization schemes. We have demonstrated that the distributed controller with deterministic quantization ensures that the system errors remain UUB, while also offering a degree of privacy protection against eavesdroppers with access only to the quantized communication signals. In contrast, the probabilistic quantization-based controller achieves asymptotic convergence in expectation and satisfies differential privacy, thereby safeguarding the system's sensitive information even in the presence of adversaries with extensive auxiliary knowledge. Furthermore, we have formulated an optimization problem to characterize the trade-off between control performance and privacy under probabilistic quantization. Simulation results validated the theoretical analysis and provided a detailed comparison between the two quantization strategies in terms of both control performance and privacy guarantees.

\bibliographystyle{IEEEtran}
\bibliography{IEEEabrv,reference}

\end{document}